\documentclass[conference]{IEEEtran}
\IEEEoverridecommandlockouts
\usepackage[linesnumbered,ruled]{algorithm2e}
\usepackage{subfig}
\usepackage{caption}
\usepackage[pdftex]{graphicx}
\usepackage{comment}
\usepackage{amsmath}
\usepackage{todonotes}
\usepackage{multirow}
\usepackage{subfloat}
\usepackage{xcolor}
\usepackage{pifont}
\newcommand{\cmark}{\ding{51}}%
\newcommand{\xmark}{\ding{55}}%
\usepackage{url}

\newcommand{\melpuf}{{MeLPUF}}
\def\BibTeX{{\rm B\kern-.05em{\sc i\kern-.025em b}\kern-.08em
    T\kern-.1667em\lower.7ex\hbox{E}\kern-.125emX}}
\begin{document}

\makeatletter % changes the catcode of @ to 11
\newcommand{\linebreakand}{%
  \end{@IEEEauthorhalign}
  \hfill\mbox{}\par
  \mbox{}\hfill\begin{@IEEEauthorhalign}
}
\makeatother % changes the catcode of @ back to 12

\title{MeLPUF: Memory-in-Logic PUF Structures for Low-Overhead IC Authentication}

\author{\IEEEauthorblockN{Christoper Vega}
\IEEEauthorblockA{\textit{Electrical and Computer Engineering}\\
\textit{University of Florida}\\
Gainesville, USA \\
c.vega@ufl.edu}
\and
\IEEEauthorblockN{Shubhra~Deb~Paul}
\IEEEauthorblockA{\textit{Electrical and Computer Engineering}\\
\textit{University of Florida}\\
Gainesville, USA \\
shubhra.paul@ufl.edu}
\and
\IEEEauthorblockN{Patanjali SLPSK}
\IEEEauthorblockA{\textit{Electrical and Computer Engineering}\\
\textit{University of Florida}\\
Gainesville, USA \\
patanjal.sristil@ufl.edu}
\and

\linebreakand % <----- NOTE HERE, breaking after the third one!

\IEEEauthorblockN{Atri Chatterjee}
\IEEEauthorblockA{\textit{Electrical and Computer Engineering}\\
\textit{University of Florida}\\
Gainesville, USA \\
a.chatterjee@ufl.edu}
\and
\IEEEauthorblockN{Swarup Bhunia}
\IEEEauthorblockA{\textit{Electrical and Computer Engineering}\\
\textit{University of Florida}\\
Gainesville, USA \\
swarup@ece.ufl.edu}
}

\maketitle

\begin{abstract}
Physically Unclonable Functions (PUFs) are used for securing electronic devices across the implementation spectrum ranging from Field Programmable Gate Array (FPGA) to system on chips (SoCs). However, existing PUF implementations often suffer from one or more significant 
deficiencies: (1) significant design overhead; (2) difficulty to configure and integrate based on application-specific requirements; (3) vulnerability to model-building attacks; and (4) spatial locality to a specific region of a chip. These factors limit their application in the authentication of designs used in diverse applications. In this work, we propose~\melpuf: Memory-in-Logic PUF; a low-overhead, distributed PUF that leverages the existing logic gates in a design to create cross-coupled inverters (\emph{i.e.}, memory cells) in a logic circuit as an entropy source. It exploits these memory cells' power-up states as the entropy source to generate device-specific unique fingerprints. A dedicated control signal governs these on-demand memory cells. They can be dispersed across the combinational logic of a design to achieve distributed authentication. They can also be synthesized with a standard logic synthesis tool to meet the target area, power, and performance constraints. We evaluate the quality of \melpuf~signatures with circuit-level simulations and experimental measurements using  FPGA silicon (TSMC 55nm process). Our analysis shows the high quality of the PUF in terms of uniqueness, randomness, and robustness while incurring modest overhead. We further demonstrate the scalability of MeLPUF by aggregating power-up states from multiple memory cells, thus creating PUF signatures or digital identifiers of varying lengths. Additionally, we suggest optimization techniques that can be leveraged to boost the performance of \melpuf~further.
\end{abstract}


% As a general rule, do not put math, special symbols or citations
% in the abstract or keywords.

	%%%%%%%%%% Introduction %%%%%%%%%%
	
\section{Introduction}
\label{sec:intro}

\begin{figure}[!tbh]
    \centering
    
             \subfloat[SRAM PUF cell.]{%
  \includegraphics[width=1in]{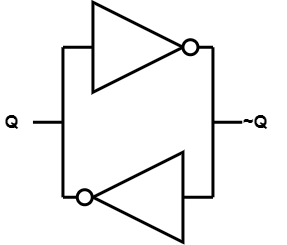}
                \label{fig:mem2}
        }
        \hfill
        \subfloat[Latch PUF cell.]{%
   \centering
    \includegraphics[width=1.2in]{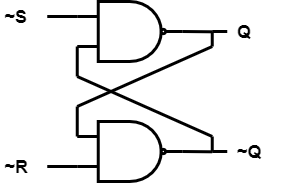}
                \label{fig:mem1}
        }
        \hfill
        \subfloat[Butterfly PUF cell.]{%
   \centering
   \includegraphics[width=1in]{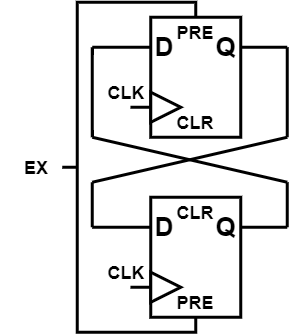}
                \label{fig:mem3}
        }
     \caption{Examples of various existing memory based PUF structures such as (a) SRAM PUF, (b) Latch PUF, and (c) Butterfly PUF.}
\label{fig:mempufs}
\end{figure}

Counterfeiting and Piracy of hardware Intellectual Property (IP) is one of the prominent trust issues faced by the semiconductor industries due to the distributed manufacturing model~\cite{hoque1, hoque2}. A robust authentication technique allows device manufacturers to detect and identify counterfeit devices, prove ownership and establish provenance. Physical Unclonable Functions (PUFs) have emerged as effective security primitives for device authentication due to their simplicity of implementation, non-resource-intensive nature, and unclonability \cite{ thesis:Ravikanth2001,thesis:Gassend_2003}. PUFs exploit the randomness introduced due to the physical variations in the manufacturing process. The designer can then use these sources of randomness to generate unique signatures. PUFs have been used as random number generators~\cite{srinipuf}, countermeasure to detect IP piracy \cite{bib:9}, and to perform chip authentication~\cite{bib:10}.

A PUF is typically characterized by a set of challenge-response pairs (CRPs). The challenges denote the input to the PUF, while the responses indicate the corresponding random output produced by the structure. Three critical features outline the performance of a PUF: uniqueness, robustness, and randomness~\cite{book:Maiti_PUFperformance}. Uniqueness signifies the ability of the PUF to produce distinct and singular responses corresponding to a set of challenges. Robustness denotes the ability to produce the same signature for the same input challenge repeatedly. Various PUF structures have been proposed in the literature that exploit both in-circuit delays~\cite{bib:ro,bib:ro2,bib:ZhangPaul2020_database_free}, and memory elements \cite{bib:weak2,bib:weak1}. 

Additionally, integrating the PUF structures into an IP introduces considerable design-level challenges. Some difficulties designers encounter include generating a reliable and unclonable signature, lowering the overheads, and ensuring that the PUF elements are resistant to tampering and modeling attacks. The security of the PUF itself is of paramount importance since the utility of the PUF structure is primarily defined by its randomness or ``uncloneability" . The uncloneablity of a PUF is its primary security measure. By revealing the secret keys or being able to replicate their behavior, one can break the main security benefit of a PUF and thus render it useless. Model building attacks~\cite{bib:weak1,bib:ai} have shown to be a significant concern in this regard as they effectively model and predict the behavior of certain PUFs. Therefore, there is a need to address and alleviate these concerns.

Over the years, various research works have focused on addressing these concerns. Delay-based PUFs broadly support a large number of challenge-response pairs but are vulnerable to model-building attacks. On the other hand, memory PUFs are harder to model synthetically, but they incur significant overheads due to their low entropy \cite{bib:PUFanalysis,bib:modeloverview}. The authors in~\cite{bib:gan} and~\cite{bib:cui} propose improvements to the delay PUFs to improve the uniqueness, reliability, and reduce the overheads. However these structures are vulnerable to model-building attacks. On the other hand, the authors in ~\cite{bib:bad} introduce a re-configurable DRAM PUF that is resistant to model-building attacks but incur a significant runtime overhead as the signature read-out time can range from 20--90s. Thus existing PUF implementations either suffer a mix of large overheads and low attack resistance or require a complicated design process to integrate the overall PUF structure into the IP.

In this work, we introduce \melpuf: \textbf{Me}mory-in-\textbf{L}ogic \textbf{PUF}; a PUF that addresses the aforementioned concerns. \melpuf~~comprises two components: a pair of cross-coupled inverters that acts as the source of entropy and control logic that facilitates the designer/user in capturing the generated PUF responses. Unlike traditional PUF structures that need to be integrated separately in a fixed region,\melpuf~structures can be integrated throughout the design in a distributed fashion without incurring overheads. We also observe that the distribution of the PUF elements through the circuit logic renders \melpuf~~ resistant to tampering attacks, thereby exhibiting exceptional robustness. To our knowledge, \melpuf~~is the first fully synthesizable PUF structure that the designer can easily integrate into the design even at the earlier stages (e.g., RTL-level) in the manufacturing process.

The significant contributions of this work are listed below:

\begin{itemize}
    \item We propose~\melpuf, a fully synthesizable novel PUF structure. \melpuf~uses the startup states of independent in-circuit memory elements as an entropy source. 
    \item The PUF structure can be integrated at higher design-abstractions, and in a distributed fashion thus incurring less overheads and enhancing the overall security. To the best of our knowledge,~\melpuf~~is the first PUF structure of this kind that can be directly integrated into digital designs at higher abstraction levels.
    \item We perform HSPICE Monte-Carlo and FPGA simulations on \melpuf~~circuit models to validate its source of entropy. The simulation results confirm a near-optimal performance of the generated signatures regarding uniqueness, robustness, and randomness. Through hardware implementation on FPGAs, we also verify that our proposed \melpuf~~structure incurs low overhead and can be integrated into various designs.
    \item Finally, we compare \melpuf~~with a few state-of-the-art PUF implementations, and we show that our proposed implementation outperforms those existing structures with high confidence.
\end{itemize}

%\vspace{-10pt}
We organize the rest of the paper as follows: We discuss the current state-of-the-art PUF implementations in Section~\ref{sec:related} and introduce the fundamental concepts of \melpuf~~in~\ref{sec:Motivation}. We then specify the procedures for integrating \melpuf~~into a given design in Section~\ref {sec:methodology}. The experimental setup and the evaluation results for simulation are discussed in Section~\ref{sec:simulation_results}, while Section~\ref{sec:results} portrays the assessment of hardware implementation and analysis of \melpuf~~as a distributed authentication mechanism.  \textcolor{black}{In Section~\ref{sec:ml_attack} we evaluate the overall security of \melpuf~by discussing the resistance of \melpuf~against model-building, tampering and side-channel attacks.} Additionally, we compare the performance of \melpuf~~with other state-of-the-art PUF implementations in Section~\ref{sec:comp}.

%%%%%%%%%% Background and Motivation  %%%%%%%%%%

\section{Background and Related Works}
\label{sec:related}

 A PUF consists of a set of challenge-response pairs (CRPs) for authentication. The challenges denote the input to the PUF, while the responses indicate the corresponding random output produced by the PUF. PUFs are classified into strong and weak PUFs based on their ability to generate a large number of random responses~\cite{bib:tax}. Strong PUFs utilize challenge-response pairs (CRP) that change the response when a new input is applied. This results in an ample challenge response space; however, they are vulnerable to model-building attacks \cite{bib:weak2,bib:weak1}. Weak PUFs, on the other hand, typically have smaller challenge response space. Typically the CRP scale linearly with each other, as such area overheads for generating substantial responses. However, this one-to-one challenge-response behavior can offer resistance to model-building attacks, such as in the case of the SRAM PUF, which makes them favorable for commercial applications \cite{bib:weak2}. However, as there is a smaller response space, the responses must possess a high degree of randomness to ensure no significant overlap in the responses generated between different devices. 
Weak PUFs are often implemented as memory PUFs. The random startup state or the result of the metastable state of the memory cell serves as the entropy source. Examples of existing memory-based PUFs are shown in Fig. \ref{fig:mempufs}.

Over time there have been several works regarding memory and FPGA implemented PUFs. The authors in~\cite{bib:gan} introduced FPGA-based RO PUF utilizing internal multiplexers of a Lookup Table (LUT), and the authors in~\cite{bib:cui} used XOR gates to reduce the area overhead. However, a significant disadvantage of RO PUFs is their vulnerability to model-building attacks \cite{bib:mlatk}. While XOR gates can increase the resistance to machine learning attacks, the improvement is not reasonably significant, and the prediction accuracy remains relatively high. Another drawback of RO PUF is the RO-locking effect. RO-locking occurs when the RF-interference bounds the RO's frequency to that of the RF wave~\cite{bib:rolock}. In \cite{bib:tero} the authors proposed a PUF exploiting the transient effect ring oscillator (TERO), which mitigates such effects.
\begin{figure*}[!t]
    \centering
    \subfloat[Unit structure of~\melpuf.]{
    \begin{minipage}[t!]{0.28\textwidth}
  \includegraphics[width=1.7in]{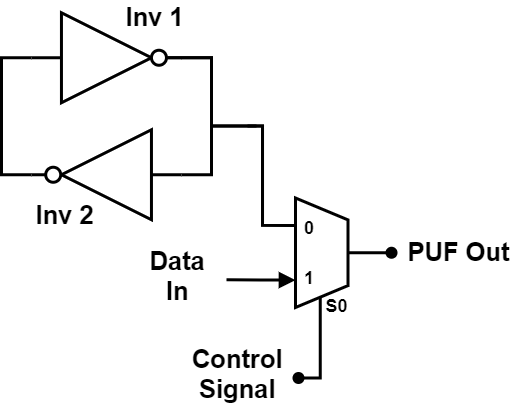}
                \label{fig:sy1}
\end{minipage}}%\hfill%
    \subfloat[\melpuf, highlighted in red, embedded in a combinational circuit.]{
    \begin{minipage}[t!]{0.34\textwidth}
   \centering
   \includegraphics[width=2.3in]{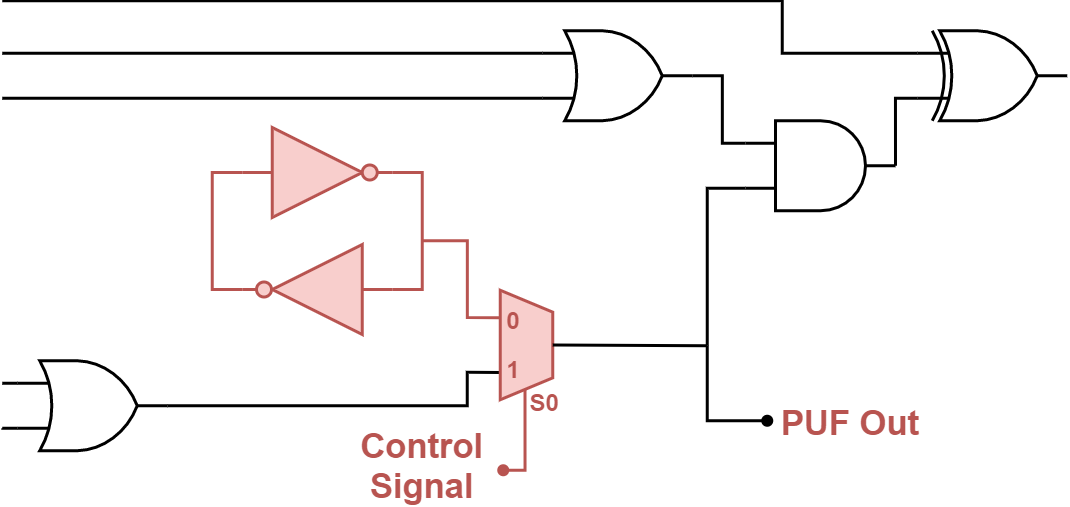}
                \label{fig:sy2}
    \end{minipage}}
    \hfill
    \subfloat[\melpuf~with the different steps of the signature generation process highlighted.]{
    \begin{minipage}[t!]{0.34\textwidth}
   \centering
   \includegraphics[width=2.3in]{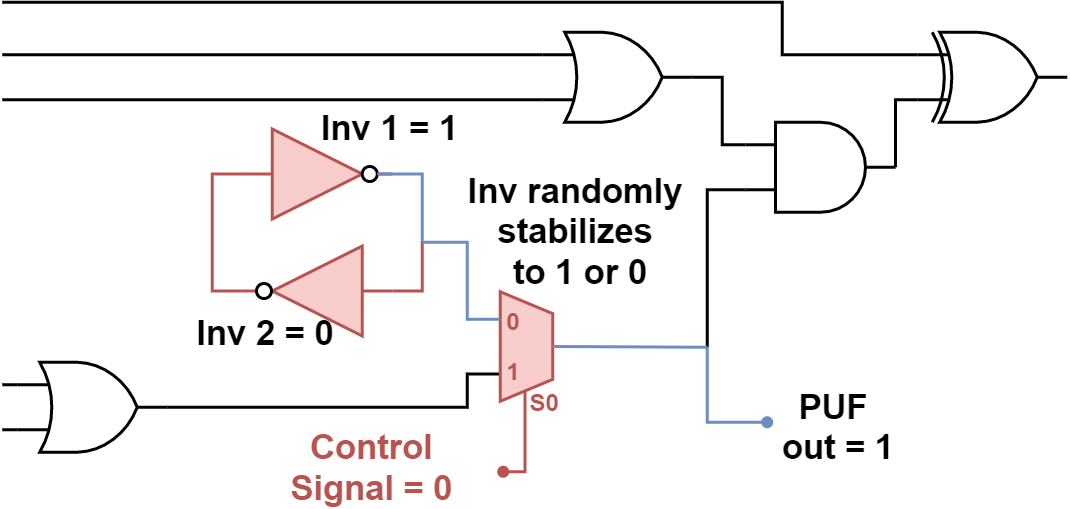}
                \label{fig:sy3}
    \end{minipage}}
    
     \caption{The structure (a), implementation (b), and working principle (c) of \melpuf. Each element of \melpuf~can be configured into a memory cell (cross-coupled inverter pair) using a control signal. The power-up states of the memory cells act as the entropy source for the PUF which can then be extracted by setting the Control Signal to 0.} \label{fig:PUFstruct}
\label{fig:overall}
\vspace{-10pt}
\end{figure*}

In \cite{bib:cui}, the authors presented a memory-based PUF using XOR gates to model latches on an FPGA. However, its comparatively lower uniqueness properties (around 40\%) make it less desirable for practical applications as a PUF. Another approach utilizes the memory write failures as a source of entropy to generate unique signatures, as shown in~\cite{bib:meca}. However, if the target design does not have a suitable memory array, it incurs additional implementation costs. The authors of ~\cite{bib:bad} proposed an interesting approach using re-configurable DRAM PUF based on refresh pausing. However, this approach requires a dedicated memory module. Additionally, the signature generation process is time-consuming, taking 20 -- 90 seconds to complete.

 Among other low-overhead memory-based PUFs, one-time programmable memory (OTPM) based weak PUFs~\cite{bib:OTPM,bib:OTPS}, flip-flop-based PUF~\cite{bib:intrinsic}, shift register-based PUFs~\cite{bib:wang, bib:ams} are significant. However, they are incompatible with all platforms (e.g., FPGA) or, the high clock frequency requirements limit their implementations.
 To reduce the overheads, designers have also used on-chip elements as sources of entropy for a PUF. The authors in~\cite{bib:scan, bib:ZhangPaul2020_database_free} invoked such an approach to utilize the boundary scan chain to generate secret keys by exploiting the path delays between scan cells. While this significantly reduces overhead, the size of the scan-chain limits the signature length. Designers also adopted similar approaches using analog elements for low-cost PUF implementations~\cite{bib:DACPUF,bib:trapdoor,bib:256,bib:paul2021}.

\section{Motivation}
\label{sec:Motivation}
Through the analysis of previous works, we can see a need for low overhead, machine learning resistant PUFs that are easily integrated into designs. These requirements pose an interesting research and design challenge. We now highlight some key points to illustrate the benefits of \melpuf~over the existing PUF structures.

\subsection{Overhead}

Existing PUF implementation often requires standalone structures. Ring Oscillator PUFs, for example, need groups of ring oscillators to generate a single bit. Moreover, SRAM PUFs require a separate SRAM module. Depending on the size of the signature, this can result in a significant area overhead for a given circuit. The insertion of the PUF elements sets \melpuf~~apart from other memory PUFs. Instead of adding a dedicated memory array (such as SRAM), which increases the overheads, \melpuf~~structures can be integrated with the combinational logic. This allows the designer to implement the PUF structures using unused logic space in the Logic Element (LE) fabric and thus reduces the overheads significantly. Moreover, the designer can integrate the PUF structures in designs that do not contain an onboard SRAM array with minimal overhead. 

\subsection{Ease of Integration}
The designer can easily integrate the \melpuf~ elements into the design fabric. As we show in subsequent Sections, the designer can integrate \melpuf~ at higher design abstractions such as RTL-level. Doing so, allows the PUF structures to blend organically into the design and significantly reduces overheads. \melpuf~~can be integrated using the conventional EDA flow without any considerable changes.

\subsection{Resistance to Model-building, Tampering and Removal attacks}

\melpuf~~ utilizes independent PUF structures that can be implemented separately, thus allowing the PUF structures to be integrated easily into existing designs. The individual PUF elements are distributed throughout the design, thus providing an inherent model-building attack resistance. The independent nature of each PUF element makes it harder for the attacker to model the overall signature, find a function that represents the relationship between the challenges and responses, or observe a correlation between different PUF element's responses \cite{bib:weak2}. Hence, performing model-building attacks on a design with \melpuf~ is complicated. Additionally, the distributed nature of the PUF element makes it harder for the attacker to perform tampering attacks. Since the PUF structure is integrated with the logic, the attacker cannot easily remove the PUF elements for performing a removal attack.

	%%%%%%%%%% Mel-PUF Methodologies  %%%%%%%%%% 

%%%%%%%%%%%%%%%%%%%%%%%%%%%%%%%%%%%%%%%%%%%%%%%%%%%%%%%%%%%%%%%%%%%%%%%%%%%%%%%%%%%%
\section{\melpuf~~Structure and Implementation } \label{sec:methodology}
In this Section, we discuss the architecture of \melpuf.  We first discuss the various steps involved in integrating \melpuf~~structures into a given design. We then discuss the various steps involved in the signature generation and extraction.

\begin{algorithm} [t]
\caption{The procedure of MeLPUF integration into a design}
\label{algo:melpuf_sig_gen}
% \LinesNumbered
% \line(1,0){250
% \begin{algorithmic}
% {\bf Algorithm 1: The procedure of PSDM} \\
\textbf{Input}: $\mathcal{D}$ Synthesized Target Design\;
\textbf{Input}: $\mathcal{T}$ Set of Target Gate types \;
\textbf{Input}: $\mathcal{P}_{Template}$ \melpuf~~template\;
\textbf{Input}: $\mathcal{N}_{PUF}$ number of PUF structures to be inserted\;
initialization\;
Let $G$[0...n] array containing candidate gates for PUF insertion\;
$k\leftarrow 0$\tcp*[l]{Number of PUF elements inserted}
$n\leftarrow 0$\tcp*[l]{Total number of candidate Gates}
\For{$gate \in$ $\mathcal{D}$ $netlist$}{
  \If{$gate~type$ matches $\mathcal{T}$ $target~gate~type$}{
  \If{$Slack_{gate}$>0}{
    $G$[$i$]$\leftarrow gate$\;
    $n \leftarrow n+1$\;
    }
   \ElseIf{$Slack_{gate}$<=0}{
    Mark $gate$ as critical\;
    }
    }
  }
\eIf{$n$ == 0 or $ n $<$ \mathcal{N}_{PUF}$}{
  Report ``Failure not enough target gates''\;
  }{
  SortSlack($G$)\tcp*[l]{high to low}
  \While{$k<\mathcal{N}_{PUF}$}{
    Insert $\mathcal{P}_{Template}$ at output of $G$[$k$]\;
    $k \leftarrow k+1$\;
    }
    Check area, power, and delay of the MeLPUF inserted netlist\;
    \textbf{Output}: MeLPUF inserted netlist, locations of inserted PUF, Report: area, power, and delay\;
    }
\end{algorithm}

\subsection{Entropy Source} \label{sec:system_arch}
\melpuf~~consists of a cross-coupled inverter pair and a control element. Upon startup, the bi-stable element's output is unknown and in a metastable state. This output is read-only and serves as the PUF response. Fig.~\ref{fig:sy1} illustrates the unit structure of the proposed \melpuf. This structure allows us to distribute and place the PUF in the datapath of a combinational circuit. The controller MUX is then used to switch between the PUF response and logic output. Thus, the circuit can operate in two modes: a) PUF mode, which enables the designer to sample the PUF responses, and b) functional mode, which transfers the logic output of the gate to its fanout cones.

Fig.~\ref{fig:sy2} depicts \melpuf~~ implementation where we deploy a bi-stable memory element in a combinational circuit. The \melpuf~~circuitry is highlighted in red, with original datapaths marked in black. During power-up, the memory cell's output results in a random state, which we use as our entropy source. Fig. \ref{fig:sy3} outlines the datapaths for generating a response from the implemented PUF. The output of one inverter is connected to a MUX, where the latter acts as the control element. When the control signal is LOW, the MUX allows the PUF signature to be captured at startup. When the control signal is HIGH, the circuit switches to functional mode and resumes regular operation.

\subsection{\melpuf~~Integration}
% \subsection{\melpuf~~Integration and Signature Extraction}
% \noindent 
We now describe the various steps involved in integrating the PUF elements into the circuit. The overall flow of the \melpuf~~template integration into a specific design is summarized in Algorithm~\ref{algo:melpuf_sig_gen}. The integration flow takes in the synthesized netlist $\mathcal{D}$, the set of target gates $\mathcal{T}$, the \melpuf~~template $\mathcal{P}_{Template}$, and the number of PUF structures that need to be inserted as inputs $\mathcal{N}_{PUF}$.  The algorithm then identifies the elements to be targeted in the design (lines 9-21). The algorithm identifies as follows: if a gate matches the target gate type ($\mathcal{T}$) (line 10), we check if the gate has sufficient timing slack for a replacement. Timing slack is defined as the difference between the required arrival time at the output of the gate and the actual arrival time at the output of the gate~\cite{bib:slack}. Thus a gate with positive slack is a suitable candidate for replacement while a gate with negative slack is unsuitable since replacing the gate with the PUF element will worsen the timing profile of the design $\mathcal{D}$. If a gate has sufficient positive slack, it is added to the set of candidate gates (lines 11-13) and a gate with negative slack is prevented from further iterations by marking it as critical (line 14). If enough candidate gates are not found the algorithm reports failure and exits (line 21). Otherwise, we sort the gates based on slack (line 23), and iteratively add the PUF template to the gate output (line 24-27). Candidate gates are selected for insertion based on criticality. The resulting feedback loop can be created without compromising the design functionality, as shown in Fig.~\ref{fig:sy2}. The final output is a PUF inserted netlist, a list of inserted PUF locations, and a report for area, power, and delay (line 28). We invoke the synthesis tool at the end of the \melpuf~ insertion step in Algorithm 1 (line 28) to obtain the area, power and delay values. We then compare the values with the values obtained for the baseline netlist to calculate the overheads. The time complexity of Algorithm is calculated as follows:
 The algorithm has three key parts i) Identification of target gates (lines 9-22), ii) Insertion of \melpuf~ elements (lines 23-27), and iii) Overhead estimation (line 28). 
 The first step involves a depth-first or breadth-first search of the entire graph to identify the target gates. The complexity would thus be $\mathcal{O}(V+E)$ where E is the number of Edges and V is the number of Vertices of the graph. Since the number of edges is approximately $V^2$. The overall complexity of target gate identification is $\mathcal{O}(V^2)$. In the second step, we do linear pass to replace the target gates with the \melpuf~ elements. This involves a single pass over all the vertices in the graph and thus incurs an overhead of $\mathcal{O}(V)$. The final overhead calculation step would involve one more depth-first or breadth-first traversal to update the timing, area, and power profile and would thus again incur a complexity of $\mathcal{O}(V^2)$. Thus the overall complexity of Algorithm 1 is given by the Equation~\ref{eqn:3}.

\begin{equation}
Time~for
\begin{cases}
Target~Gate~Identification +  \\ 
Insertion~of~MeLPUF~elements +  \\ 
Overhead~estimation \\ 
\end{cases}
\label{eqn:1}
\end{equation}
 \begin{equation}
     Time~Complexity=\mathcal{O}(V^2)+\mathcal{O}(V)+\mathcal{O}(V^2)
     \label{eqn:2}
 \end{equation}
 \begin{equation}
     Time~Complexity=\mathcal{O}(V^2)
     \label{eqn:3}
 \end{equation}

\subsection{Signature Generation and Extraction}
The PUF structure can be inserted and distributed throughout a circuit for use in device authentication. The bi-stable element enters a metastable state on startup, similar to a memory element. The state in which the bi-stable element stabilizes is unknown and random. The state which is settled on in then utilized as a unique signature than can be used for authentication of a device. Fig. \ref{fig:sy3} shows the steps for signature generation and extraction. When the PUF signature is read out, the control signal goes low. This switches from the functional circuit logic to the PUF output. Once this is performed, the PUF value is ready to be read out through the PUF output line and sampled. By combining the outputs of multiple PUF elements, a unique signature for the device can be generated. By comparing the signature collected from the device with that of previously sampled golden response the authenticity of the device can be confirmed. 

\subsection{Post-Synthesis Verification}
Post-synthesis verification can be handled much in the same way before the PUF elements are inserted. The control element can be set to disable the PUF and restore the circuit's logical function. The circuit can then be tested and used for verification as well as standard circuit functionality.\\

% \end{comment}
%%%%%%%%%%%%%%%%%%%%%%%%%%%%%%%%%%%%%%%%%%%%%

	%%%%%%%%%% Mel-PUF Synthesis  %%%%%%%%%%
%	\input{sections/Synthesis}

%%%%%%%%%%%%%%%%%%%%%%%%%%%%%%%%%%%%%%%%%%%%%
\section{\melpuf~Simulation and Analysis}
\label{sec:simulation_results}

A typical PUF implementation is expected to demonstrate the following three properties: uniqueness, robustness, and randomness~\cite{bib:proper}. We assess our proposed PUF structure concerning these attributes using simulation and by implementing it on an FPGA. We discuss the simulation results in this section and the hardware results in the subsequent sections.

\subsection{Setup of Simulation Environment} \label{sec:melpuf_simulation}

Physical unclonable functions exploit the randomness or entropy of physical characteristics from manufacturing process variations. As a result of the fluctuations of parameters such as threshold voltage, transistor width or length, gate delay, \emph{etc.}, the power-up values of circuit elements get affected, and \melpuf~takes advantage of it. For this purpose, we create a transistor-level model of the \melpuf~circuit in HSPICE using the $45$ nm high-performance CMOS process node from Predictive Technology Model (PTM). We use an input clock frequency of 100 MHz signal to this circuit along with the nominal supply voltage, $V_{DD}=1.0$ V. We initialize all the circuit nodes as `0' at the beginning of the simulation. We consider the manufacturing process variations by combining the effects of physical specifications such as $t_{ox}$, $W$, $L$,  \emph{etc.} into a single parameter: the threshold voltage, $V_{th}$~\cite{Mukhopadhyay_modeling_failure}. We model these process variations as a function of  $V_{th}$ by using a Gaussian distribution for 10,000 \melpuf~circuit models/instances with $\sigma_{inter-die}=25\%$. We apply the in-built  parameter distribution function of HSPICE, \textit{AGAUSS}, to shift/skew the transistor threshold values to generate this distribution as depicted in Fig.~\ref{fig:simulation_vth}. To characterize the effect of manufacturing process variations, we perform Monte-Carlo simulation on power-up values at nominal temperature, $T_{NOM}=25$ $^\circ$C on the generated model. Each Monte-Carlo simulation run results in a 1-bit power-up value for that specific \melpuf~circuit instance, and we use that as a corresponding binary signature. Fig.~\ref{fig:simulation_colormap} illustrates the gray colormap of collected power-up values for 10,000 \melpuf~instances, where the dark dots indicate power-up states of `1' and white dots represent the `0's. Our experiment discovered that 5033 out of the total 10,000 simulated circuit instances startup with  `1', and the rest appear to have a bias towards `0' as their power-up values. To investigate the performance as a PUF, we build a circuit model with 64 instances of \melpuf~for each Monte-Carlo simulation to collect 64-bit PUF responses from each run; thus, we collect 640,000 power-up values/signatures from all 10,000 runs. We evaluate the security performance of the simulated \melpuf~responses in terms of uniqueness, robustness, and randomness.

\subsection{Simulation Results} 
\label{sec:results_simulation}

\begin{figure} [!tbh]
        \centering
         \subfloat[ Distribution of transistor $V_{th}$ variations for Monte-Carlo simulation.]
        {                \includegraphics[width=2in]{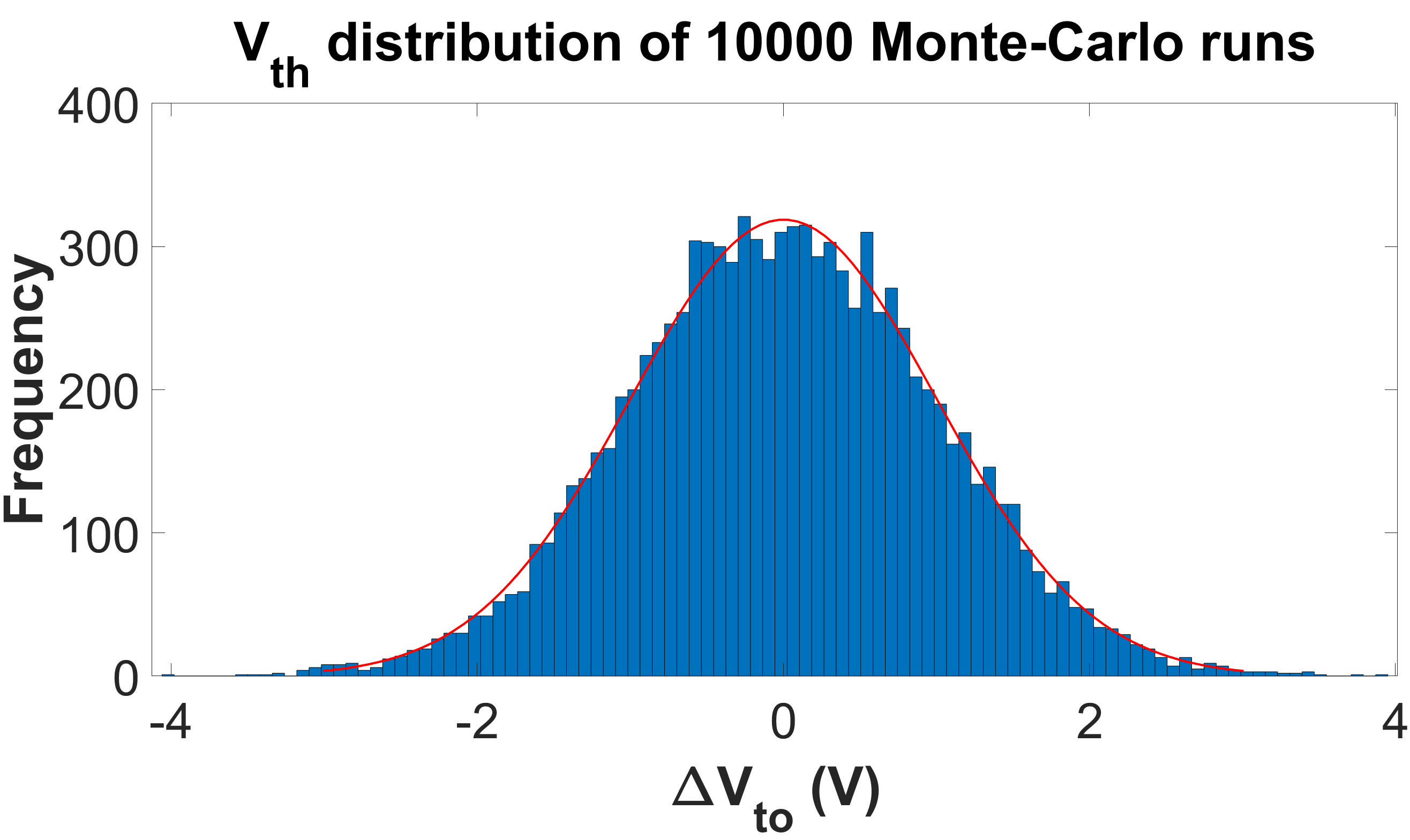}
                \label{fig:simulation_vth}
        }
        \hfill
         \subfloat[Gray colormap of simulated \melpuf~power-up values.]
        {               \includegraphics[width=2in]{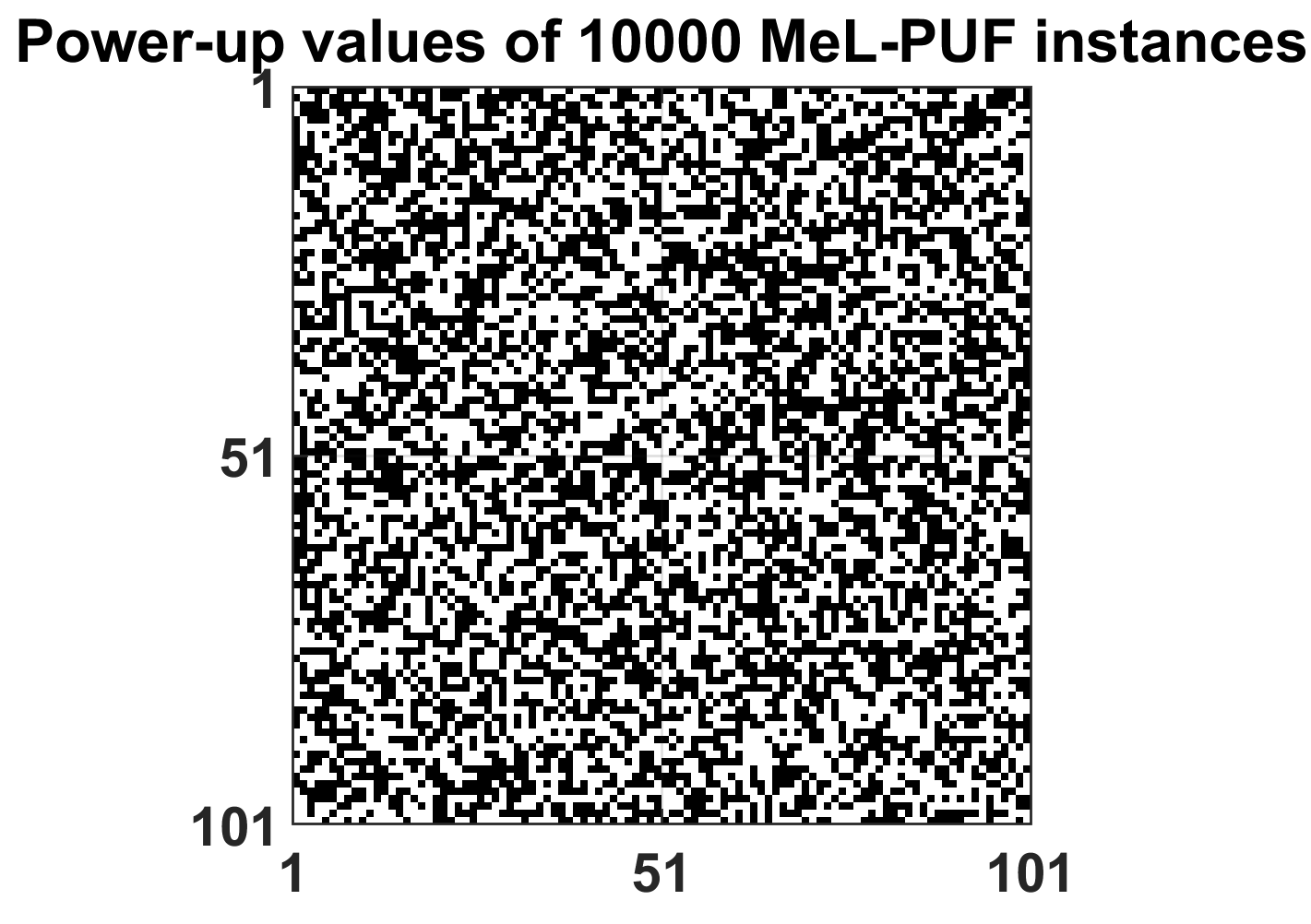}
                \label{fig:simulation_colormap}
        }
        \centering
        \caption{HSPICE simulation setup and results for Monte-Carlo experiments. Figure (a) shows the distribution of the $V_{th}$ values used to collect the 640,000 \melpuf~signatures and (b) shows the distribution of the collected powerup states. Dark dots indicate a '1' and white space represents a '0'.} \label{fig:simulation_results}
\vspace{-10pt}
\end{figure}  

\begin{figure} [!tbh]
        \centering
        \subfloat[Uniqueness (Inter-HD) results for simulated \melpuf.]
        {                \includegraphics[width=2in]{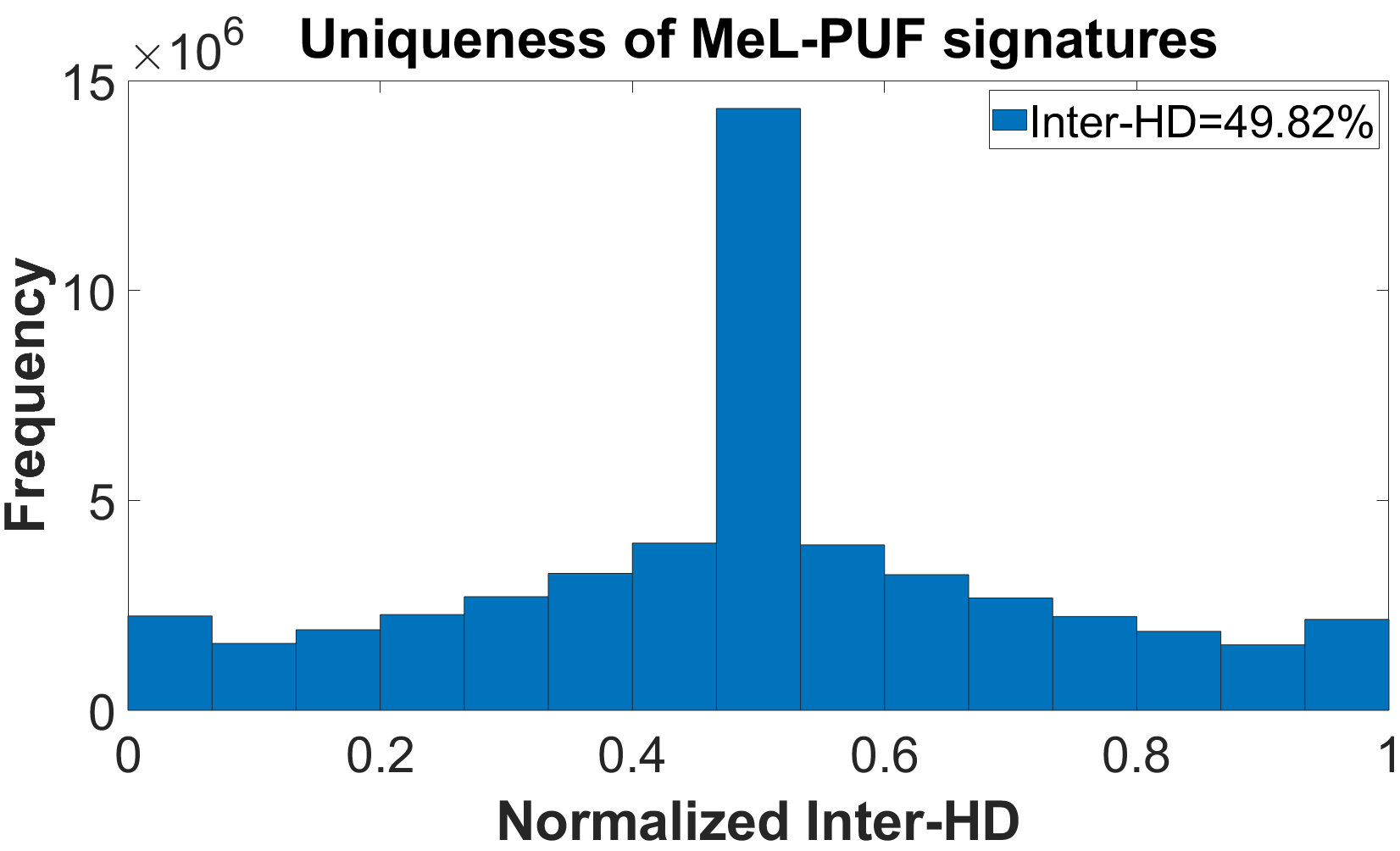}
                \label{fig:simulation_uniqueness}
        }
        \hfill
        \subfloat[\melpuf~robustness comparison at different temperature levels.]
        {                \includegraphics[width=2in]{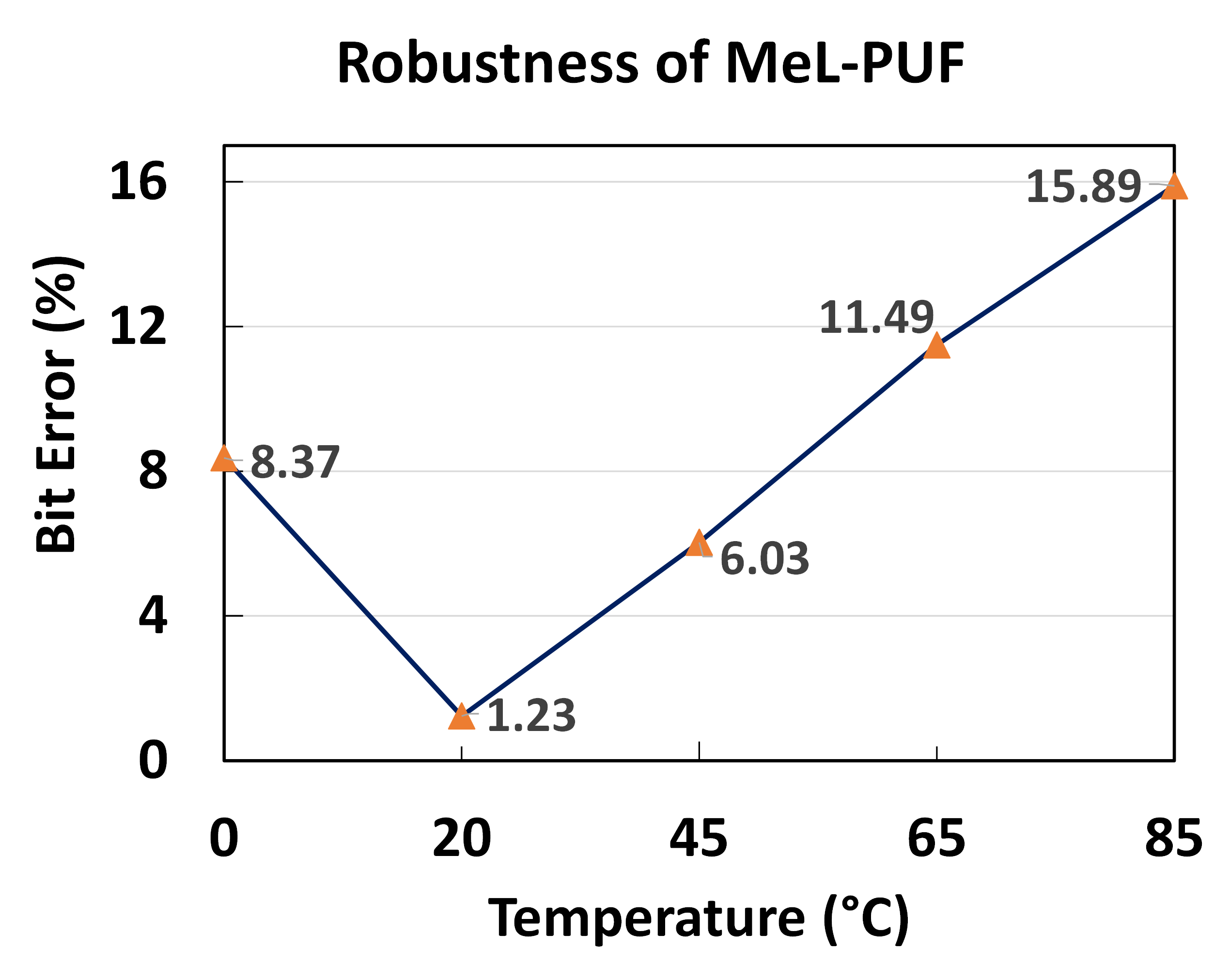}
                \label{fig:simulation_robustness_comparison}
        }
        \centering
        \caption{The Uniqueness~(a) and Robustness~(b) plots obtained using the HSPICE simulation setup for a 64-bit response size.} \label{fig:simulation_results}
\vspace{-10pt}
\end{figure}  

\subsubsection{Uniqueness Analysis} 
\label{sec:uniqueness_simulation}
We assess the uniqueness of \melpuf~ by calculating the average inter-chip Hamming distance (inter-HD). Inter-HD is the difference in the number of bits obtained by applying the same challenge to two separate PUF entities. It is calculated using Eqn.~\eqref{equ:inter}~\cite{book:Maiti_PUFperformance}.

\begin{equation}
  \text{Inter-HD}_{avg}=  \frac{2}{M(M-1))}\sum_{u=1}^{M-1}\sum_{v=u+1}^{M}\frac{HD(S_{u},S_{v})}{n}\times 100\%,
    \label{equ:inter}
 \end{equation}
 where $S_u$ and $S_v$ are  $n$-bit response vectors from $u^{th}$ and $v^{th}$ chip ($u\neq v$) for the same challenge, and $M$ is the total number of chips.
 
Fig.~\ref{fig:simulation_uniqueness} illustrates the uniqueness results of the collected signatures (at $T_{NOM}=25$ $^\circ$C) in terms of  inter-HD. Here, the X-axis shows the percentage of bits difference between the responses, and the Y-axis shows the number of times the bit difference occurs. We observe that the simulation design of \melpuf~ demonstrates $49.82\%$ inter-HD, which is close to an ideal case.

\subsubsection{Robustness Analysis} \label{sec:robustness_simulation}
We use the average intra-Hamming distance (intra-HD) as defined in Eqn.~\eqref{equ:intra} for evaluating the robustness of \melpuf~\cite{book:Maiti_PUFperformance}. 
\begin{equation}
\text{Intra-HD}_{avg}=\frac{1}{N}\sum_{u=1}^{N}\frac{HD(S_{u, 1},S_{u,2})}{n}\times 100\%,
\label{equ:intra}
\end{equation}
where $S_{u,1}$ and $S_{u,2}$ are the $n$-bit response vectors from the $u^{th}$ chip due to the challenge over 1\textsuperscript{st} and 2\textsuperscript{nd} measurement, respectively for a total number of $N$ chips.

Intra-HD is the difference in response offered when two different challenges are applied to the same PUF implementation. It is estimated using Eqn.~\eqref{equ:intra}. We assess the intra-HD by simulating the operation of \melpuf~over five different temperature levels: 0$^\circ$, 20$^\circ$, 45$^\circ$, 65$^\circ$, and 85 $^\circ$C (at nominal $V_{DD}=1.0$ V). Fig.~\ref{fig:simulation_robustness_comparison} shows the percentage of bit-errors in terms of average intra-HD at different temperature levels. We observe that the PUF exhibits $8.37\%$, $1.23\%$, $6.03\%$, $11.49\%$, and $15.89\%$ bit-errors 0$^\circ$, 20$^\circ$, 45$^\circ$, 65$^\circ$, and 85 $^\circ$C respectively. %This indicates that our PUF may show less robustness when operating at more extreme temperatures. Temperatures below 65 $^\circ$C show good robustness. 

%\textcolor{black}{Does that mean it is good or bad?}

\subsubsection{Randomness Analysis} \label{sec:randomness_simulation}
Besides uniqueness and robustness, we also investigate the randomness property of the implemented PUF model. Randomness is one of the most critical security properties of a PUF because it determines whether an adversary can predict the PUF responses during any attack scenario. To evaluate the randomness, we perform a simulation on our \melpuf~model and collect one million response bits to assess them using the NIST (National Institute of Standards and Technology) randomness test suite~\cite{bib:nist_randomness}. Table~\ref{tab:nist_simulation} summarizes the results of 14 randomness tests from the suite that we perform at a significance level, $\alpha=0.001$. We explore that the generated sequences successfully pass all of the performed tests at a \textit{p}-value higher than the defined significance level. Thus we conclude that the simulated \melpuf~responses are random with a confidence level of 99.9\%.

%%%%%%%%%%%%%%%%%%%%%%%%%%%%%%%%%%%%%%%%%%%%%%%%%%%%%%%%%%%%%%%%%%%%%%%%%
% Table generated by Excel2LaTeX from sheet 'Sheet1'
\begin{table}[!t]
  \centering
  \caption{NIST Test Suite results and corresponding \textit{p-value}s of one million simulated \melpuf~response bits using a significance level $\alpha=0.001$.} \label{tab:nist_simulation}
%   \resizebox{\columnwidth}{!}{
    \setlength{\arrayrulewidth}{1pt}
    \begin{tabular}{|l|c|c|}
    \hline 
    \textbf{Statistical Test} & \textbf{\textit{p-value}}  & \textbf{Pass?} \\[.5ex] \hline
    Frequency & 0.002043   & Y \\[.5ex] \hline
    Block Frequency & 0.935716   & Y \\[.5ex] \hline
    Cumulative Sums (Forward) & 0.013569  & Y \\[.5ex] \hline
    Cumulative Sums (Backward) & 0.002374   & Y \\[.5ex] \hline
    Runs  & 0.455937   & Y \\[.5ex] \hline
    Longest Run & 0.574903   & Y \\[.5ex] \hline
    FFT   & 0.085587   & Y \\[.5ex] \hline
    Non-overlapping Template & 0.911413  & Y \\[.5ex] \hline
    Overlapping Template & 0.001399  & Y \\[.50ex] \hline
    Approximate Entropy  & 0.275709   & Y \\[.5ex] \hline %($m=7$)
    Serial (Forward) & 0.122325  & Y \\[.5ex] \hline
    Serial (Backward) & 0.066882  & Y \\[.5ex] \hline
    Linear Complexity & 0.213309  & Y \\[.5ex] \hline
    Rank & 0.075719  & Y \\[.5ex] \hline
    \end{tabular}%
%   }
\vspace{-10pt}
\end{table}%
%%%%%%%%%%%%%%%%%%%%%%%%%%%%%%%%%%%%%%%%%%%%%%%%%%%%%%%%%%%%%%%%%%%%%%%%%
	%%%%%%%%%% Results and Analysis  %%%%%%%%%%
	\section{Implementation and Analysis}
\label{sec:results}

The following section details the evaluation of \melpuf~hardware implementation and the analysis of experiment results. While the simulation results give promising quality of PUF signatures, they may not reflect real-world manufacturing variations, affecting the accuracy of the results. Therefore, it is necessary to perform a hardware evaluation of \melpuf~to investigate the behavior in a real-world scenario.

\subsection{Experimental Setup}

\begin{figure*} [!tbh]
        \centering
        
        {
                \includegraphics[width=3in]{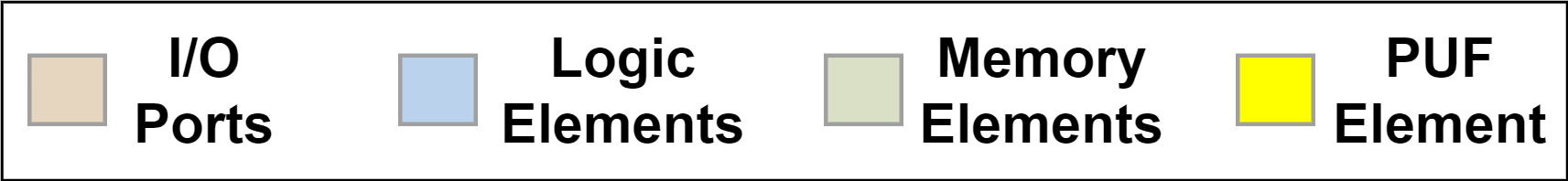}
                \label{fig:llllll}
        }
        
         \subfloat[D1: PUF structure clustered around a single area.]
        {
                \includegraphics[width=1.4in]{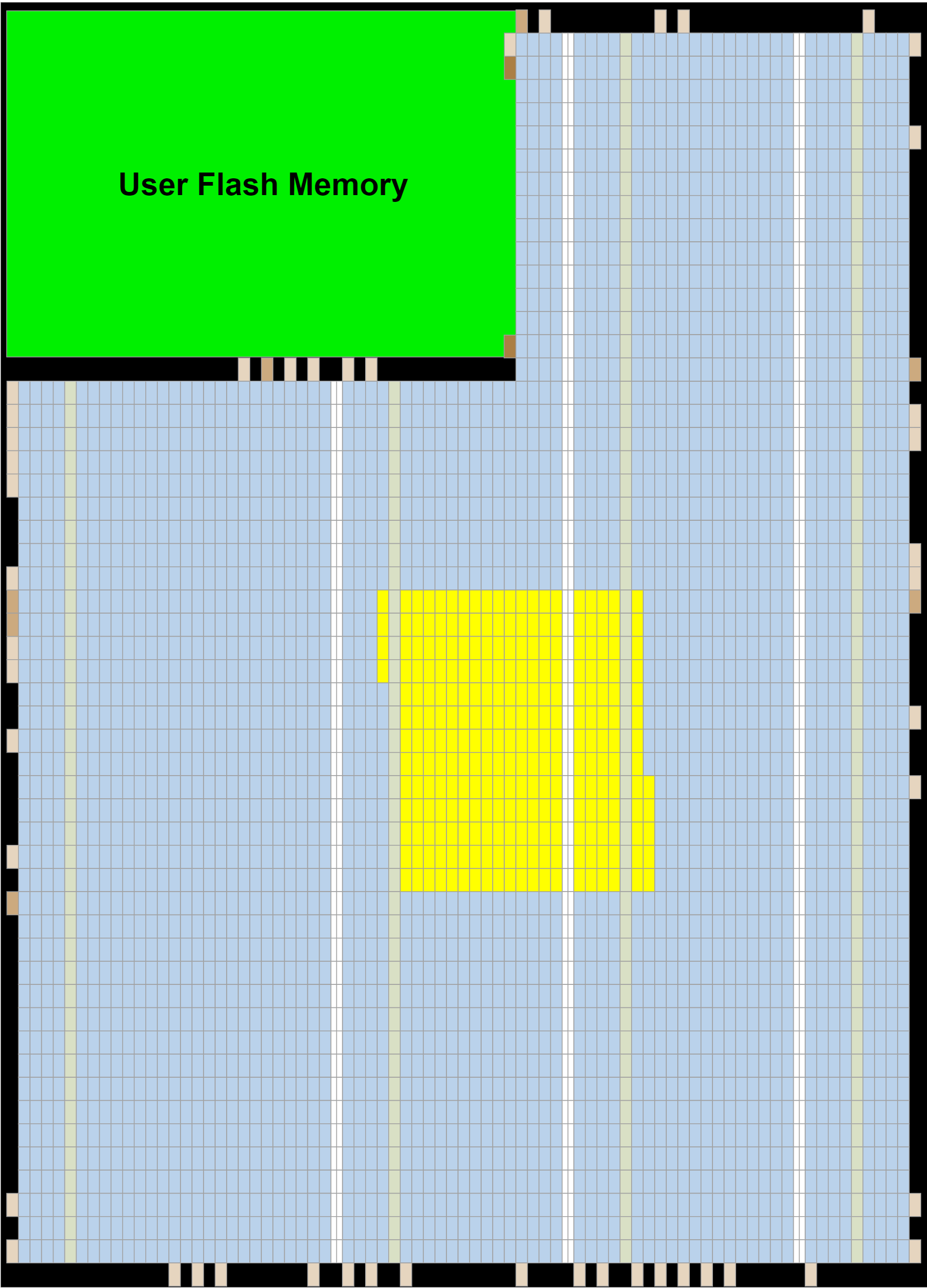}
                \label{fig:h3}
        }
        \hfill
        \subfloat[D2: PUF distributed as $32\times32$-bit groups.]
        {
                \includegraphics[width=1.4in]{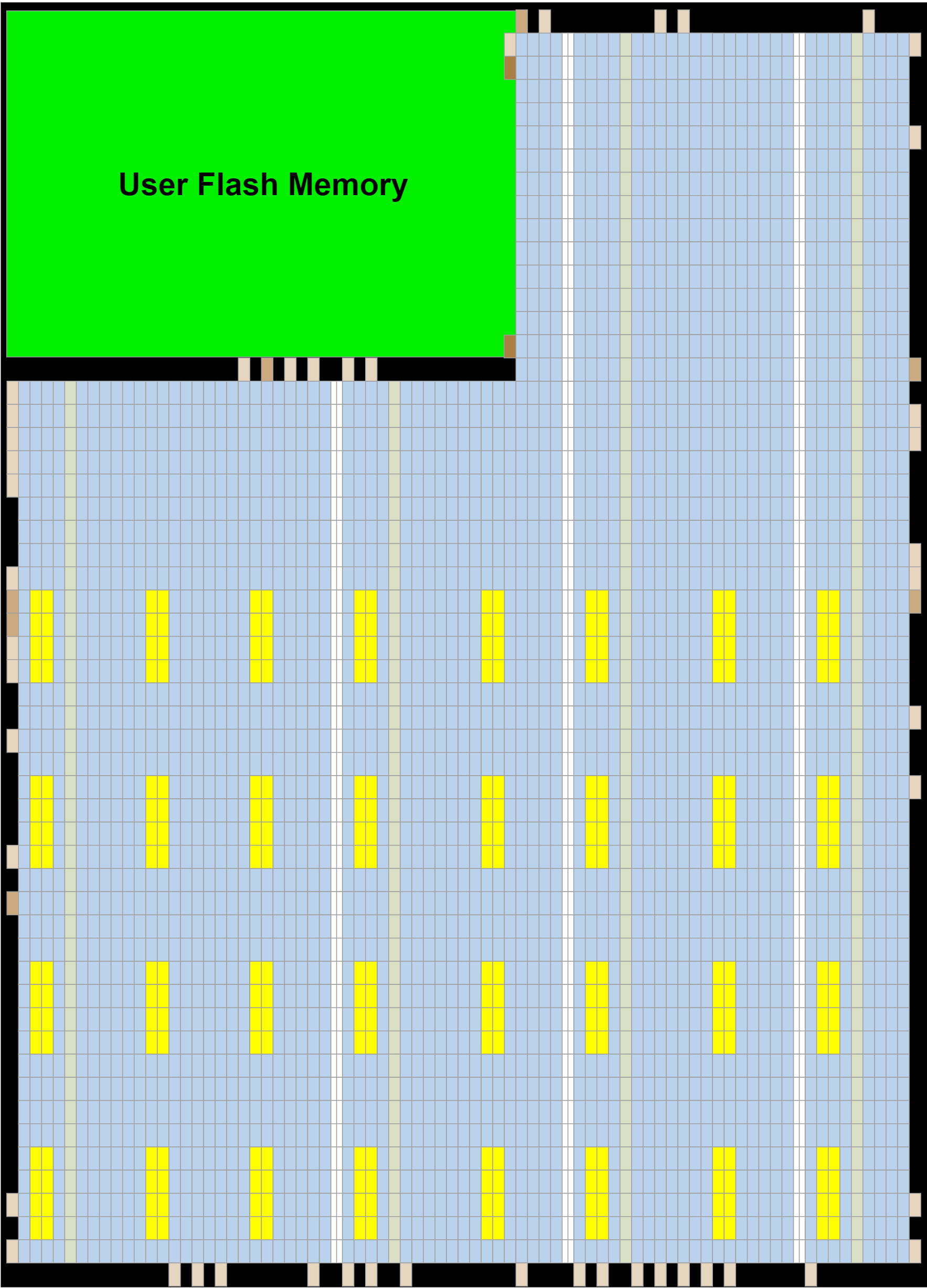}
                \label{fig:h7}
        }
        \hfill
        \subfloat[D3: PUF distributed as $64\times16$-bit groups with adjacent LABs.]
        {
                \includegraphics[width=1.4in]{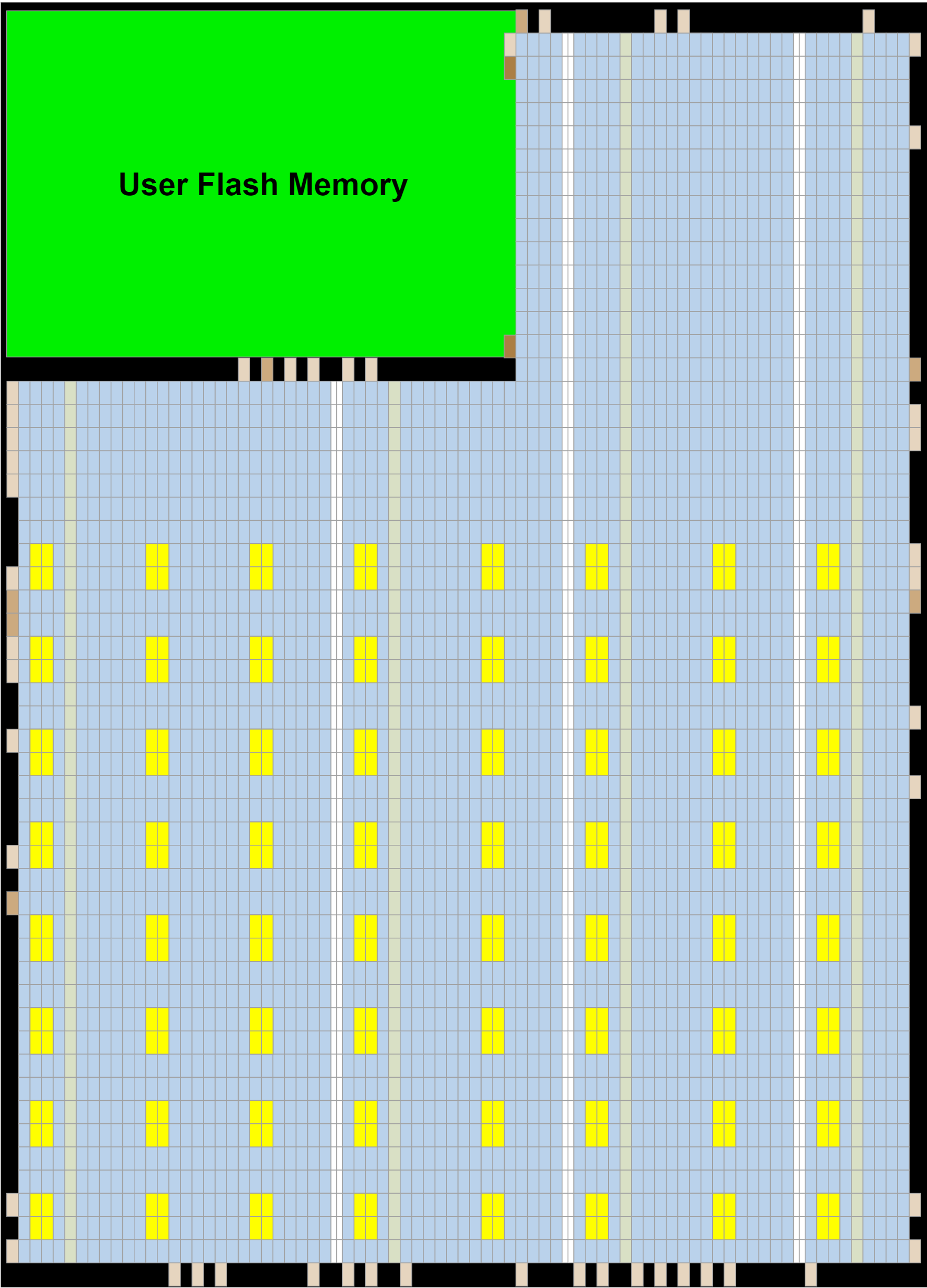}
                \label{fig:h8}
        }
        \hfill
        \subfloat[D4: PUF distributed as $64\times16$-bit groups with no adjacent LABs.]
        {
                \includegraphics[width=1.4in]{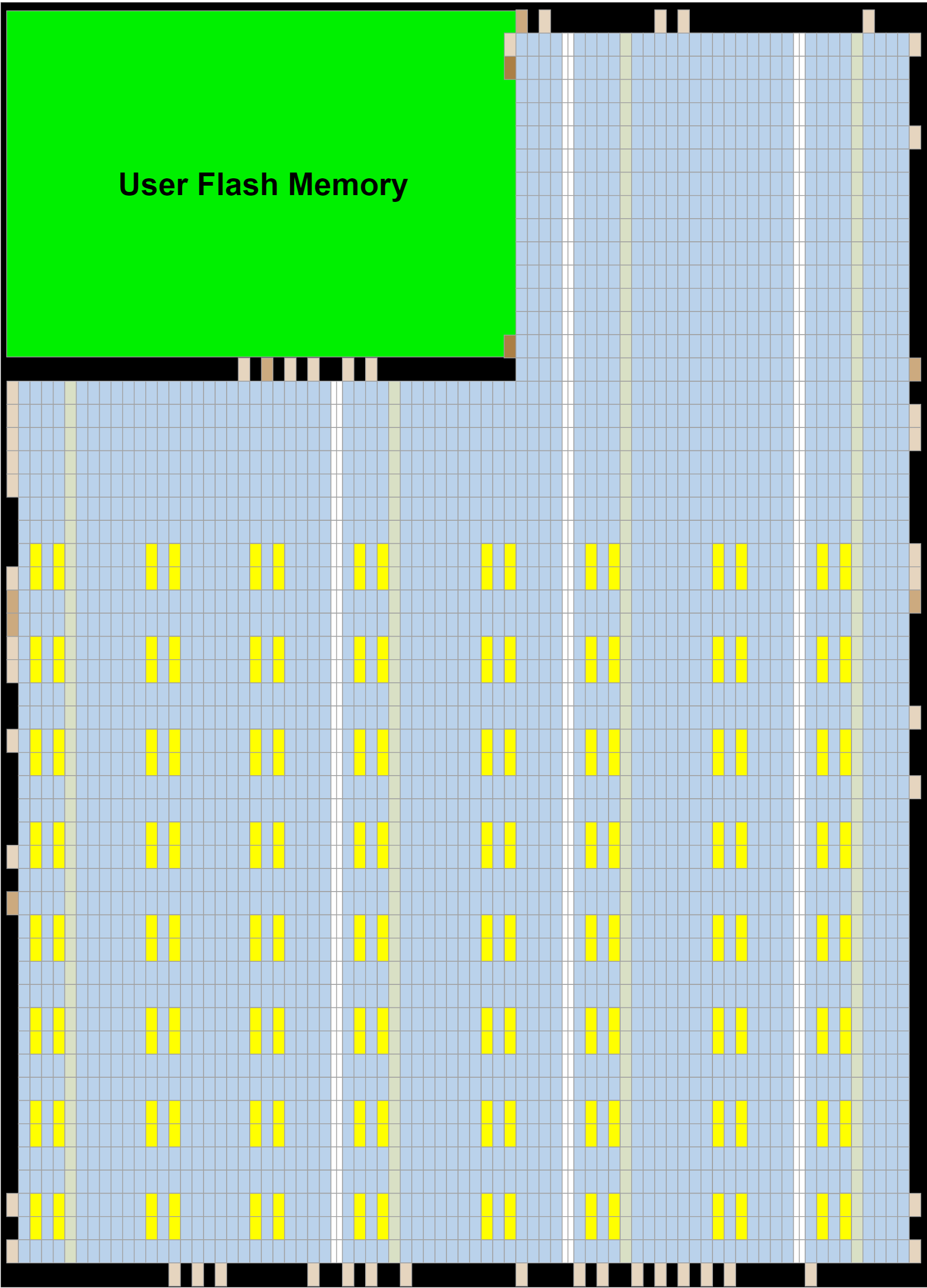}
                \label{fig:h9}
        }
        \caption{FPGA floorplans with both clustered and distributed MeLPUF~layouts. We perform experimental measurements with four different levels of the spatial distribution of PUF elements to investigate the effect of PUF cell placements.} \label{fig:heatmaps}
\vspace{-10pt}
\end{figure*} 

% \begin{figure*}[t]
%     \centering
%   \includegraphics[width=0.9\textwidth]{fig/updated/mappings_2crop.png}
%     \caption{Quartus Prime views of PUF circuit after compilation. (a) shows the PUF as shown by the netlist viewer; and (b) shows the chip-level mapping from the Chip Planner. Teal boxes show the logic elements used for PUF circuitry while Blue ones show circuit logic.}
%     \label{fig:qmap}
%     % \vspace{10pt}
% \end{figure*}

We validate our \melpuf~structure by implementing it on an Intel MAX 10 FPGA (10M50SAE144C8G) based custom platform known as HaHa board~\cite{bib:yang, bib:yang_shubhra2021_HaHa}. We use the ISCAS85 benchmark circuits to insert the \melpuf~structure while ensuring that the LUTs (Lookup Tables) are not simplified or combined with any other logic \cite{bib:ICAS}. The PUF output is routed to RAM (Random Access Memory) and used to extract the signature through Intel Quartus Prime software's in-system memory content editor. We place the PUF LUTs using the assignment editor to test different FPGA regions. We ensure that the PUF structures are mapped to the same LUT locations to ensure a repeatable signature generation for robustness and uniqueness analysis.
We sample four different LUT distribution regions for each board and collect four signatures from each region. We use these regions to calculate the generated signature's uniqueness, robustness, and randomness. We collect the signatures under nominal conditions (3.3 V at 25 \textdegree C). Additionally, we sample the signatures by varying the supply voltages within the range of 3.3 V -- 2.0 V to quantify the robustness of our designed PUF.

\subsection{Experimental Results}
We generate four different PUF configurations D1-D4, as shown in Fig.~\ref{fig:heatmaps}. These configurations are used to inspect the PUF element's distribution effect on signature quality. D1 structure clusters the PUF elements adjacent to one another. D2 and D3 disperse the PUF through the board in 32$\times$32 and 64$\times$16 clusters, respectively, isolating PUF elements from one another. D4 is another 64$\times$16 distribution while separating horizontally adjacent elements. We sample the signatures obtained from these regions and explore that the difference between signatures from regions D1, D2, and D3 is virtually indistinguishable. However, D4 results in an enhancement to the quality of the signature. We further investigate the reason behind this enhancement in uniqueness. We observe that when adjacent LEs are used (e.g., D1, D2, and D3), there is a tendency for the elements to obtain the same value. One reason for this can be due to systemic variation ~\cite{bib:sysvar}. Elements located next to each other have a possibility of having similar variations due to the manufacturing process. Another reason is that routing may become unbalanced when mapped tightly, resulting in a biased output. By taking this into account, we can improve the uniqueness of \melpuf~implementation by spacing out elements horizontally and vertically, such as in D4. For the following experiments the placement D4 is used for collection of experimental results.

\begin{figure} [!tbh]
        \centering
         \subfloat[Histogram depicting the inter-HD values for the experimental measurements, indicating uniqueness.]
        {
                \includegraphics[width=2in]{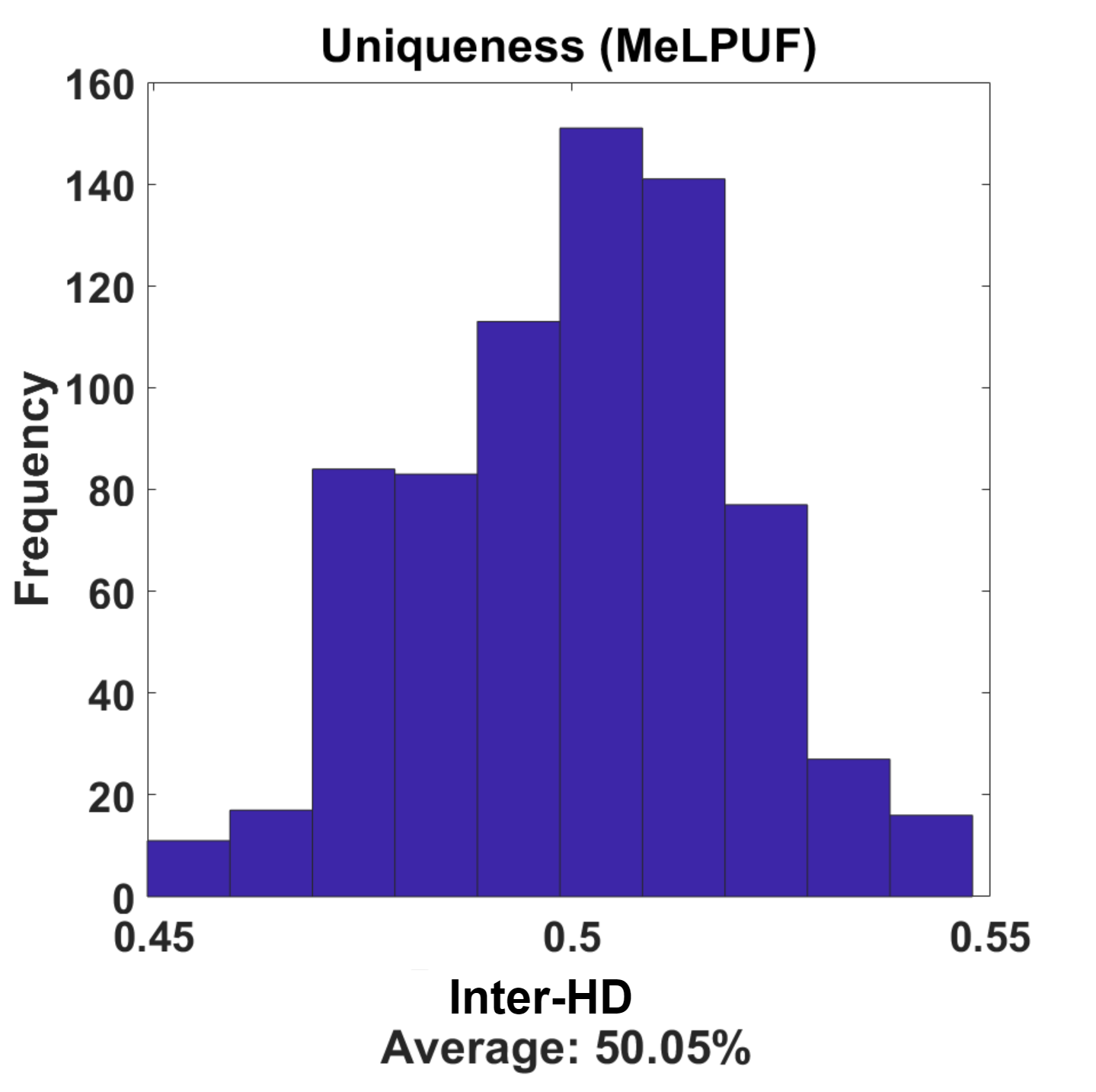}
                \label{fig:inter}
        }
        \hfill
        \subfloat[Histogram depicting the  intra-HD values for the experimental measurements, indicating robustness.]
        {
                \includegraphics[width=2in]{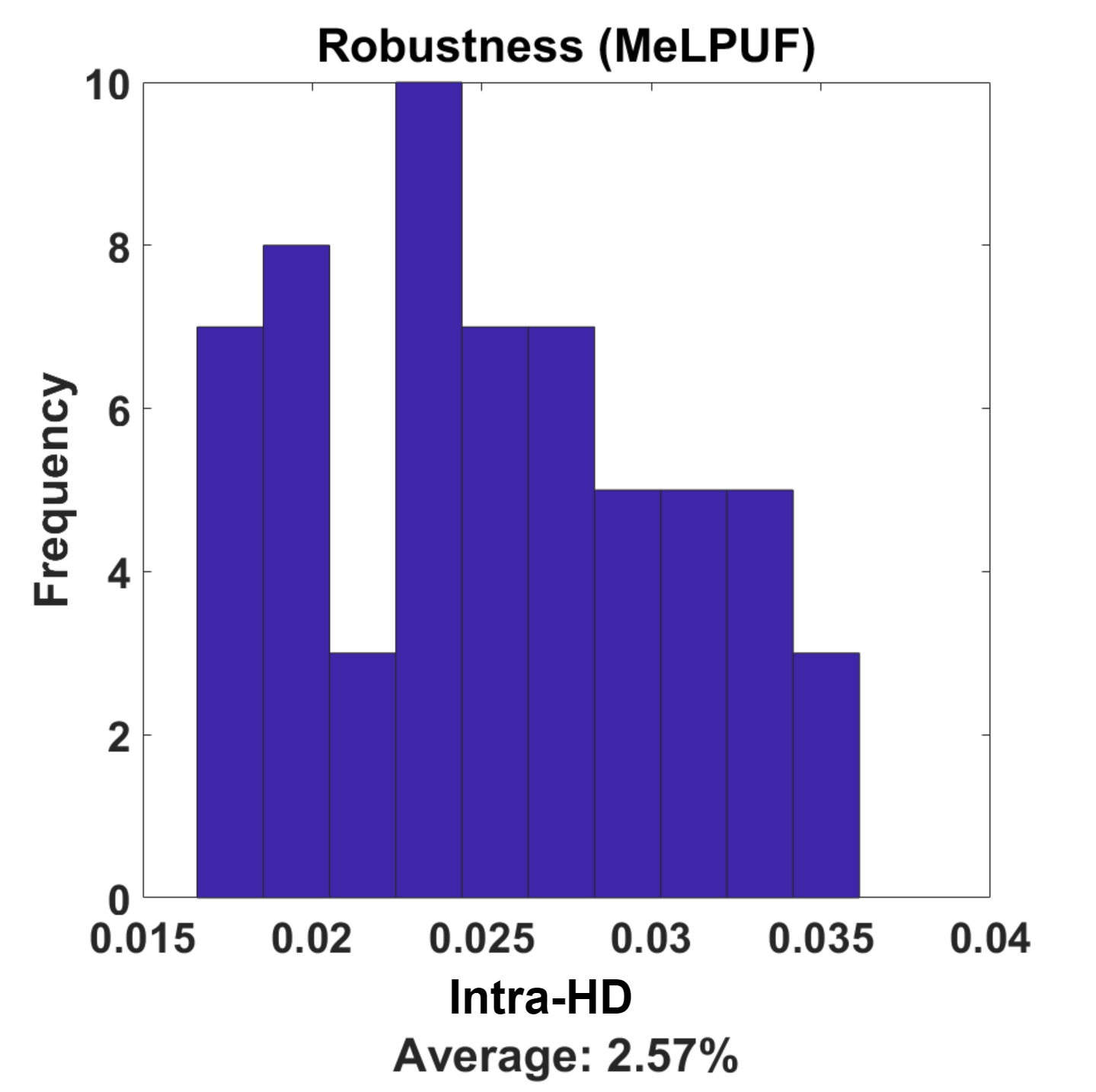}
                \label{fig:intra}
        }
        \centering
       \caption{The Uniqueness~(a) and Robustness~(b) results obtained using the hardware evaluation on an FPGA. We use the D4 configuration to obtain the results.} \label{fig:resutsPUF}
\vspace{-10pt}
\end{figure} 

%%%%%%%%%%%%%%%%
% \begin{comment}

\begin{table*}[!tbh]
\centering
\scriptsize
 \caption{NIST Test Suite results for \melpuf~signatures generated from hardware implementation. Signatures from ten different boards (i.e., 10 different chips) are collected for each region. The results indicate the number of signatures are above the threshold for the passing \textit{p}-value ($\alpha= $ 0.001) and the uniformity of the computed \textit{p}-values. The suite automatically runs the Cumulative Sums tests in two opposite directions (forward and backward) of the signature, indicated by $S_1$ and $S_2$. }
\begin{tabular}{|c|c|c|c|c|c|c|c|c|}
\hline
\multirow{5}{0.5cm}{\textbf{Test}} & \multicolumn{2}{c|}{\textbf{D1}} & \multicolumn{2}{c|}{\textbf{D2}} & \multicolumn{2}{c|}{\textbf{D3}} & \multicolumn{2}{c|}{\textbf{D4}} \\ \cline{2-9} 
 & \multicolumn{1}{c|}{\begin{tabular}[c]{@{}c@{}}No.of\\ Passing\\ Inputs\end{tabular}} & \multicolumn{1}{c|}{\textit{p-value}}& \multicolumn{1}{c|}{\begin{tabular}[c]{@{}c@{}}No.of\\ Passing\\ Inputs\end{tabular}} & \multicolumn{1}{c|}{\textit{p-value}} & \multicolumn{1}{c|}{\begin{tabular}[c]{@{}c@{}}No.of\\ Passing\\ Inputs\end{tabular}} & \multicolumn{1}{c|}{\textit{p-value}} & \multicolumn{1}{c|}{\begin{tabular}[c]{@{}c@{}}No.of\\ Passing\\Inputs\end{tabular}} & \multicolumn{1}{c|}{\textit{p-value}} \\ \hline
 \textbf{Frequency} & 8/10 &0.00000& 7/10 &0.00000& 7/10 &0.00000& 10/10 &0.73992\\ \hline
\textbf{Block Frequency} & 10/10 & 0.35049 & 10/10 & 0.00204 & 10/10 & 0.00020 & 10/10 & 0.35049 \\ \hline
\textbf{Cumulative Sums, $S_1$} & 8/10 & 0.00000 & 7/10 & 0.00000 & 10/10 & 0.00000 & 10/10 & 0.91141 \\ \hline
\textbf{Cumulative Sums, $S_2$} & 6/10 & 0.00000 & 7/10 & 0.00000 & 10/10 & 0.00000 & 10/10 & 0.73992 \\ \hline
\textbf{Runs} & 8/10 & 0.00000 & 6/10 & 0.00000 & 6/10 & 0.00000 & 10/10 & 0.03517 \\ \hline
\textbf{Longest Run} & 9/10 & 0.01791 & 9/10 & 0.00888 & 10/10 & 0.21331 & 10/10 & 0.73992 \\ \hline
\textbf{Rank} & 10/10 & 0.00000 & 10/10 & 0.00004 & 10/10 & 0.00095 & 10/10 & 0.00020 \\ \hline
\textbf{FFT} & 10/10 & 0.00888 & 10/10 & 0.01791 & 10/10 & 0.06688 & 10/10 & 0.21331 \\ \hline
\end{tabular}

 \label{tab:NISTinit}
 \vspace{-10pt}
\end{table*}
% \end{comment}
%%%%%%%%%%%%

\subsubsection{Uniqueness Analysis}

\begin{figure}[!tbh]
    \centering
    
             \subfloat[The structure of \melpuf~ as implemented with the balancing registers. The bistable loop is highlighted in blue.]
        {
   \includegraphics[width=0.35\columnwidth]{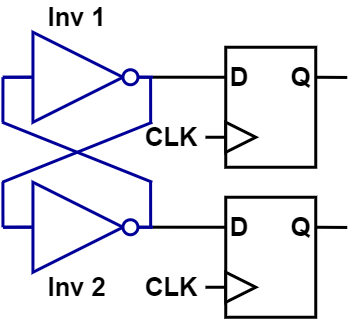}
                    \label{fig:regbalance}
        }
        \hfill
        \subfloat[LUT configuration of the \melpuf~elements on the MAX10 FPGA. Blue indicates an inverter mapped into an LUT.]
        {
                \includegraphics[width=1.4in]{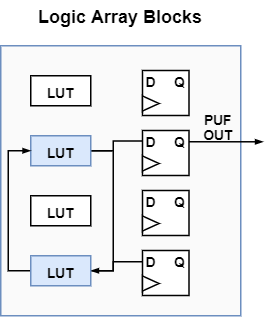}
                \label{fig:LUTimp}
        }
       \caption{The hardware implementation of \melpuf~on the MAX10 FPGA, which shows the path balancing using registers and the mapping for LUT and register elements.} \label{fig:implementation}
    \vspace{-10pt}
\end{figure}

We have to consider balancing the inverter and MUX paths to achieve ideal results. Directly routing the PUFs to external connections which connect to the MUX results in additional delay causing another source of bias~\cite{bib:routing, bib:placement}. This delay is due to one inverter driving the MUX path while the other is not, adding additional delay to the driving inverter. Using the register in each logic element (LE) as a buffer, we could balance the connection of the two inverters. Both inverters drive a register, while only one register is needed to drive the MUX indirectly. This configuration reduces the bias significantly and improves the uniqueness and robustness of the signature~\cite{bib:regiso}. An example of the connections is shown in Fig. \ref{fig:regbalance} with Fig. \ref{fig:LUTimp} showing the routing between LUTs used for PUF elements.

Fig.~\ref{fig:inter} demonstrates the uniqueness histogram between the tested chips. The X-axis shows the percentage of bit differences between signatures. The Y-axis represents the frequency at which these differences occur. The computed average inter-chip Hamming distance is 50.05\%, giving us close to the ideal value (50\%). 

\subsubsection{Robustness Analysis}
We assess the robustness of \melpuf~by comparing signatures from the same LUT region over multiple measurements. We explore that several LUTs exhibit different levels of stability due to manufacturing variations. Fig.~\ref{fig:intra} shows a histogram of robustness, giving our solution an average intra-chip Hamming distance is 2.57\%.

\subsubsection{Randomness Analysis}

The randomness evaluation of the experimental results are shown in Table \ref{tab:NISTinit}. The NIST test suite can give us an estimation of the probability that our PUF response is random, as the test can not confirm true randomness. We obtained signatures from ten chips and used these signatures to determine the probability of randomness of \melpuf. We use the NIST Test Suite~\cite{bib:nist_randomness} for our assessment and perform seven tests on each sequence. We compare the results collected for the four distributions D1, D2, D3, and D4 in order to determine if the placement of PUF elements can have an effect on the randomness of the signature. We observed that the placement D4, which consists of widely distributed PUF elements, is the most optimal configuration. 
%%%%%%%%%%%%%%%%%
\subsubsection{Reliability Analysis}

As a measure of PUF reliability, supply voltage variations are also considered. In this case, the typical voltage supplied to the board is 3.3 V. We collect signatures for voltages between 3.0 V -- 2.0 V and compare the signature to those of 3.3 V. The differences in the average Hamming distances are presented in Fig. \ref{tab:volt}. \textcolor{black}{The difference is calculated as the inter-HD values between PUF responses for different voltages for the same chip. We also perform the same experiments on an SRAM PUF as a comparison. }
%In this case, the difference inter-HD should ideally be 0\% as the generated signatures at different supply voltages are expected to remain the same. Significant changes only begin occurring at 2.2 V when the inter-HD increases to 4\%, with the highest being 13\% at 1.97 V, the lowest the board could function. Beyond 2.2 V, the variations were around 2.5\% which is reasonable. 
\textcolor{black}{We observe that the robustness of the PUF is not significantly affected at different voltages; thus, the hardware implementation of \melpuf demonstrates excellent performance in terms of reliability. Additionally, \melpuf~demonstrates excellent reliability to voltage variations than the compared SRAM PUF.}

% \subsubsection{Reliability Analysis}
%%%%%%%%%%%%%%%%%
\begin{comment}
\begin{table}[!tbh]
\scriptsize
\caption{Average differences in Hamming distances between previous measurements for a range of supply voltages (at $T_{NOM} = 25$  $^\circ$C) for \melpuf~ and an SRAM PUF. The minimal disparities of the computed values between two sets of measurements confirm a strong reliability feature.}
\label{tab:volt}
\centering
\begin{tabular}{|l|l|l|l|l|l|}
\hline
\textbf{Supply Voltage}                          & \textbf{3.0 V}  & \textbf{2.65 V} & \textbf{2.5 V}  & \textbf{2.2 V} & \textbf{2.0 V} \\ \hline
 \textbf{Diff. (MeLPUF)} & 2.4\% & 2.3\% & 2.4\% & 3.1\%  & 14.7\%  \\ \hline
\textbf{Diff. (SRAM) } & 8.1\% & 8.1\% & 8\% & 13.2\%  & 13.6\% \\ \hline
\end{tabular}
% \vspace{-10pt}
\end{table}
\end{comment}

\begin{figure}[!tbh]
    \centering
   \includegraphics[width=1\columnwidth]{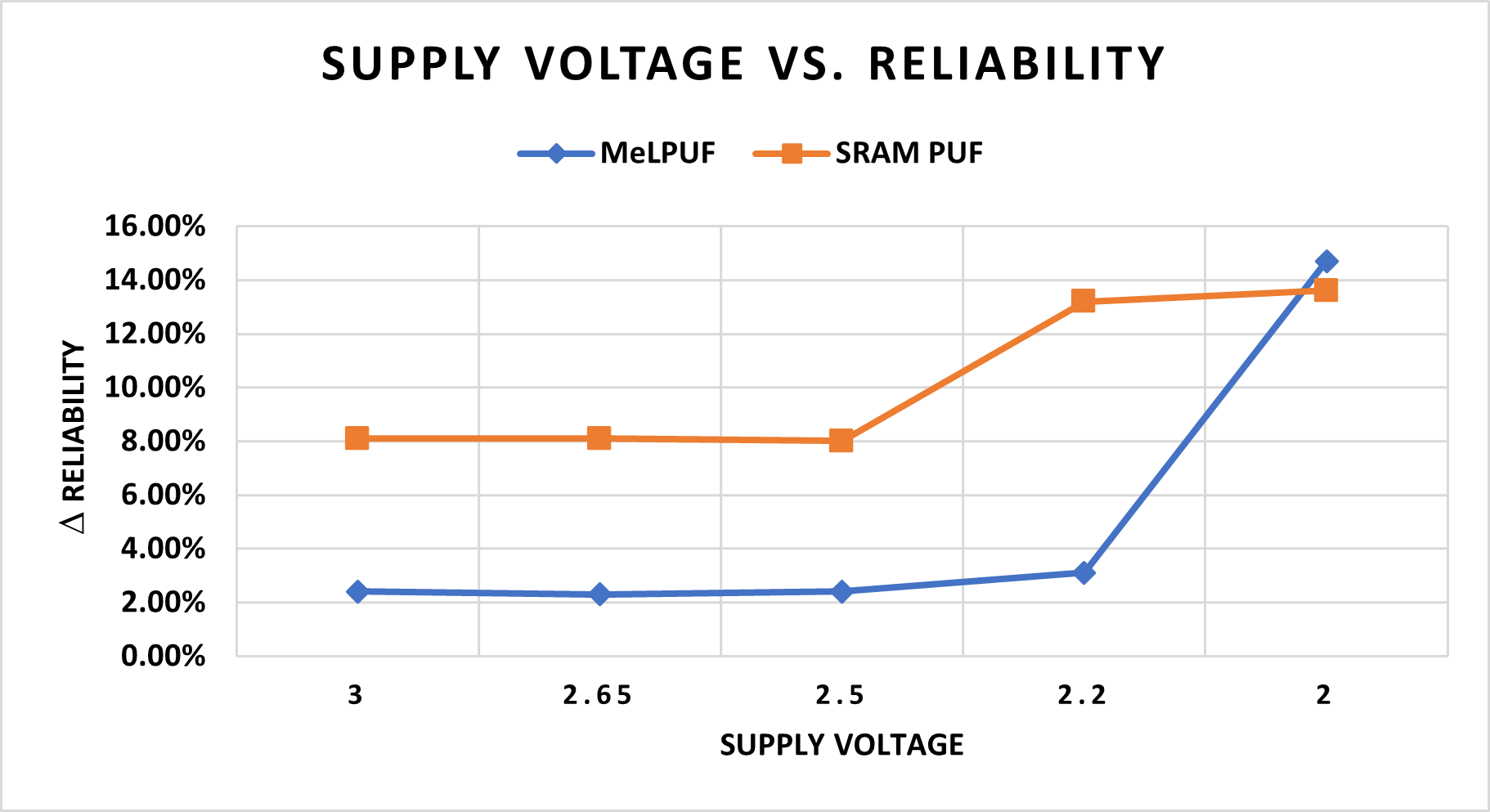}
    \caption{Average differences in Hamming distances between previous measurements for a range of supply voltages (at $T_{NOM} = 25$  $^\circ$C) for \melpuf~ and an SRAM PUF. The minimal disparities of the computed values between two sets of measurements confirm a strong reliability feature. SRAM PUF accuracy under voltage variations are obtained by measurements of SRAM blocks in the embedded hard processor in MAX10 FPGA.}
    \label{tab:volt}
    \vspace{-10pt}
\end{figure}

%%%%%%%%%%%%%%%%%%%%%%%%%%%%%%%%%%%%%%
% \begin{comment}

% \end{comment}
%%%%%%%%%%%%%%%%%%%%%%%

%%%%%%%%%%%%%%%%%%%%%%%%%%%%%%%%%%%%%%
\begin{comment}

\begin{figure*} [t]
        \centering
        
        {
                \includegraphics[width=3in]{fig/updated/ledgend.png}
                \label{fig:llllll}
        }
        
         \subfloat[D1: PUF structure clustered around a single area.]
        {
                \includegraphics[width=1.6in]{fig/updated/H3.png}
                \label{fig:h3}
        }
        \hfill
        \subfloat[D2: PUF distributed as $32\times32$-bit groups.]
        {
                \includegraphics[width=1.6in]{fig/updated/H7.png}
                \label{fig:h7}
        }
        \hfill
        \subfloat[D3: PUF distributed as $64\times16$-bit groups with adjacent LABs.]
        {
                \includegraphics[width=1.6in]{fig/updated/H8.png}
                \label{fig:h8}
        }
        \hfill
        \subfloat[D4: PUF distributed as $64\times16$-bit groups with no adjacent LABs.]
        {
                \includegraphics[width=1.6in]{fig/updated/H9.png}
                \label{fig:h9}
        }
        \caption{FPGA floor-plans with both clustered and distributed MeLPUF~layouts. We perform experimental measurements with four different levels of the spatial distribution of the PUF elements to investigate the effect of PUF cell placements.} \label{fig:heatmaps}
\vspace{-10pt}
\end{figure*} 

\end{comment}
%%%%%%%%%%%%%%%%%%%%%%%

		%%%%%%%%%% Comparison  %%%%%%%%%%
	\section{Security Analysis of~\melpuf~}\label{sec:ml_attack}

In this Section, we analyze the resistance of \melpuf~against tampering, model-building, and side-channel attacks. We discuss the effect of model-building attacks and show that it is not possible to successfully learn the behavior of \melpuf. We then evaluate the feasibility of side-channel attacks such as power, and EM. Finally, we discuss the data-corruption or the Remanence decay attack.

Typically, a model-building attack attempts to find a function that represents the relationship between the challenges and responses~\cite{bib:ai}. A modeling-building attack is typically not possible for weak PUFs~\cite{bib:weak2} because the challenge-space is not well-defined for the responses. A similar observation holds true for \melpuf~~ as the responses are not generated based on any specified input challenge. However, side-channel attacks do still present a risk and can be combined with machine learning algorithms in attempting to model the PUF~\cite{bib:ai}. 

Several different side-channel attacks have demonstrated success at cloning PUFs \cite{bib:pufem}. As such, we investigate the possibility of attacks such as Differential Power Analysis (DPA)~\cite{bib:introDPA}, Electro-Magnetic (EM) attacks~\cite{bib:EMattck,bib:EMattck2}, and Remanence Decay (RD) attacks~\cite{bib:REMattck}. Both DPA and EM side-channel attacks work on a similar principle of exploiting the electrical currents of a circuit to reveal information about the processes being performed ~\cite{bib:introDPA, bib:EMattck}. Essentially, the electrical current is changed by performing different operations on a circuit, revealing information about the processes. For example, it can be used to attack a PUF with CRPs. When a challenge is entered, it can indicate that the response is a '1' or high if the power increases. However, one countermeasure against or way to make such attacks more difficult is through balancing. It involves balancing the amount of power used or the impedance of the signal in order to make it less dependent on the data of the operations ~\cite{bib:EMattck2, bib:introDPA}. As mentioned previously, one way of balancing the signal paths for \melpuf~~is by placing a register at the output of each inverter. While it is utilized to balance the signal path delay, it can also balance the power used. As each PUF element contains two registers, this means that two opposing values represent PUF value, either a  `01' or a `10' ~\cite{bib:SCHreduction}. Thus, the power levels generated by the PUF elements can be considered balanced, making it more difficult for an attacker to determine its value. Moreover, since the inputs do not control the PUF responses, the attacker would also need to determine the order of the values extracted through DPA. Therefore, it would still be infeasible to obtain the signature.

%  \textcolor{black}{One attack that has shown promise on similar PUF structures is the RD attack. The attacker first overwrites the SRAM cells with a known value. Then they perform a fault injection attack by controlling the power-up times to change the reminance decay ~\cite{bib:REMattck}. This process causes some SRAM cells to retain their value or revert to the PUF value, allowing the attacker to determine the PUF value. This attack has been successfully performed on SRAM cells with a similar structure to \melpuf. However, one significant difference prevents an attacker from performing this attack: they do not have direct control over the PUF elements in \melpuf. This difference prevents an attacker from overwriting the PUF element with a known value, which is required to perform this attack.}

\textcolor{black}{One attack that has shown promise on similar PUF structures is the RD attack ~\cite{bib:REMattck}. The attacker first overwrites the SRAM cells with a known value. Then they perform a fault injection attack by controlling the power-up times to change the remanence decay. It causes some SRAM cells to retain their value or revert to the PUF value, allowing the attacker to determine the PUF value of the cells. This attack has been successfully performed on SRAM cells with a similar structure to  \melpuf. However, one significant difference prevents an attacker from performing this attack: they do not have direct control over the PUF elements in \melpuf. The ~\melpuf~ elements can not be written to, and it is configured only for reading out. This difference prevents an attacker from overwriting the PUF element with a known value, which is required to perform this attack.}

%https://in.nau.edu/wp-content/uploads/sites/223/2019/11/PUF-designed-with-Resistive-RAM-and-Ternary-States-1.pdf
%https://eprint.iacr.org/2015/148.pdf
%https://link.springer.com/content/pdf/10.1007/s13389-011-0006-y.pdf
%https://eprint.iacr.org/2018/620.pdf
%https://www.esat.kuleuven.be/cosic/publications/thesis-182.pdf

%Randomness results can be seen in table \ref{tab:NIST}. While the distribution of the PUF improved the uniformity of p-values for most tests, the runs tests continue to fail. This can indicate an unseen dependency between signatures that adjacent LABs do not influence. 
%Table \ref{tab:NISTinit} shows the results generated for the distribution used in figure \ref{fig:h9} running the NIST test suit. 

\begin{figure*} [t]
        \centering
         \subfloat[Comparison between the uniqueness measurements for \melpuf, SRAM PUF, and RO PUF.]
        {
                \includegraphics[width=0.25\textwidth]{fig/impv_reg/Inter_10.png}
                \label{fig:melinter}
                \includegraphics[width=0.25\textwidth]{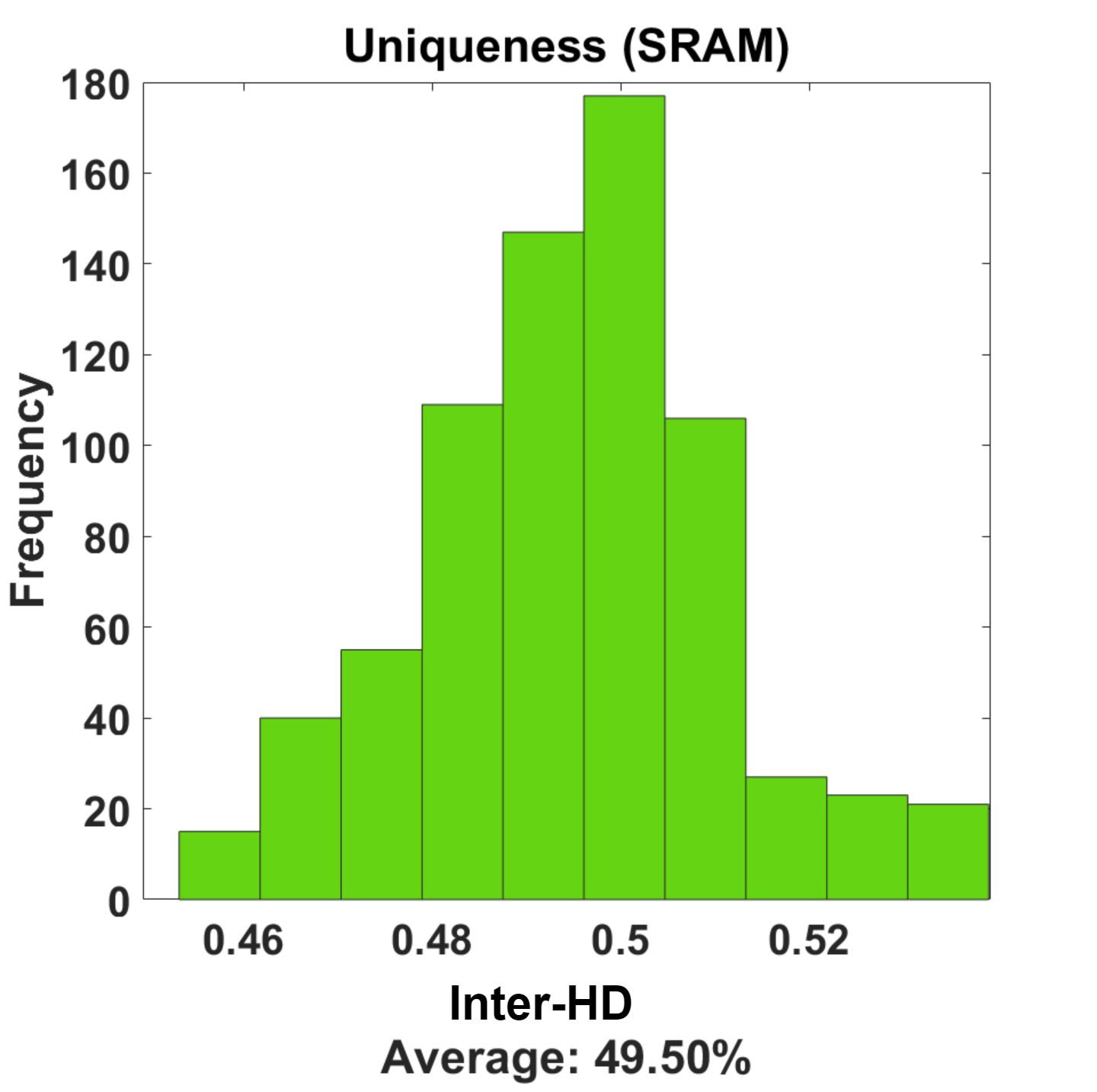}
                \label{fig:sraminter}
                \includegraphics[width=0.25\textwidth]{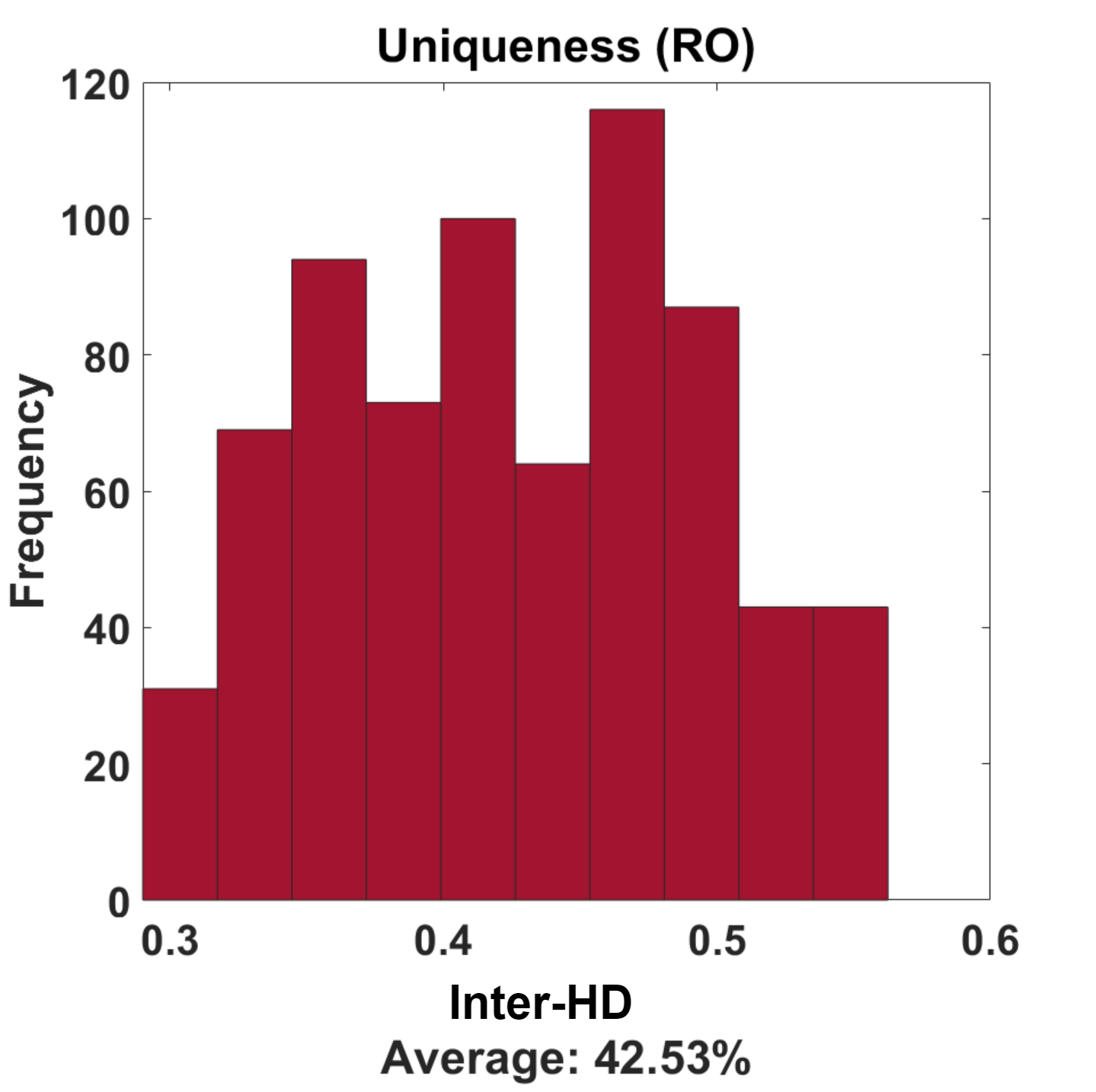}
                \label{fig:rointer}
        }
        \hfill
                 \subfloat[Comparison between the robustness measurements for \melpuf, SRAM PUF, and RO PUF.]
        {
                \includegraphics[width=0.25\textwidth]{fig/impv_reg/intra_10.png}
                \label{fig:melintra}
                \includegraphics[width=0.25\textwidth]{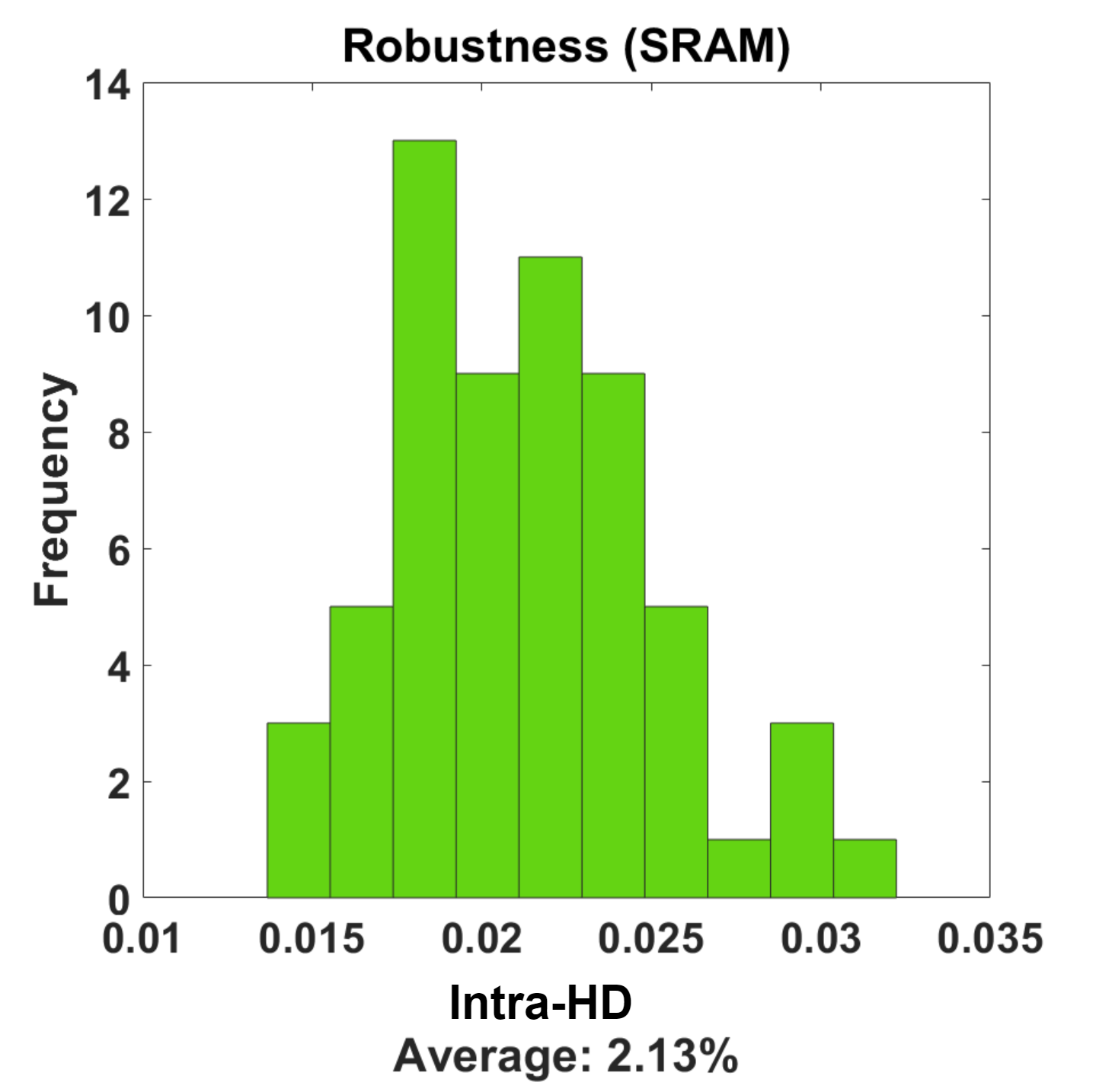}
                \label{fig:sramintra}
                \includegraphics[width=0.25\textwidth]{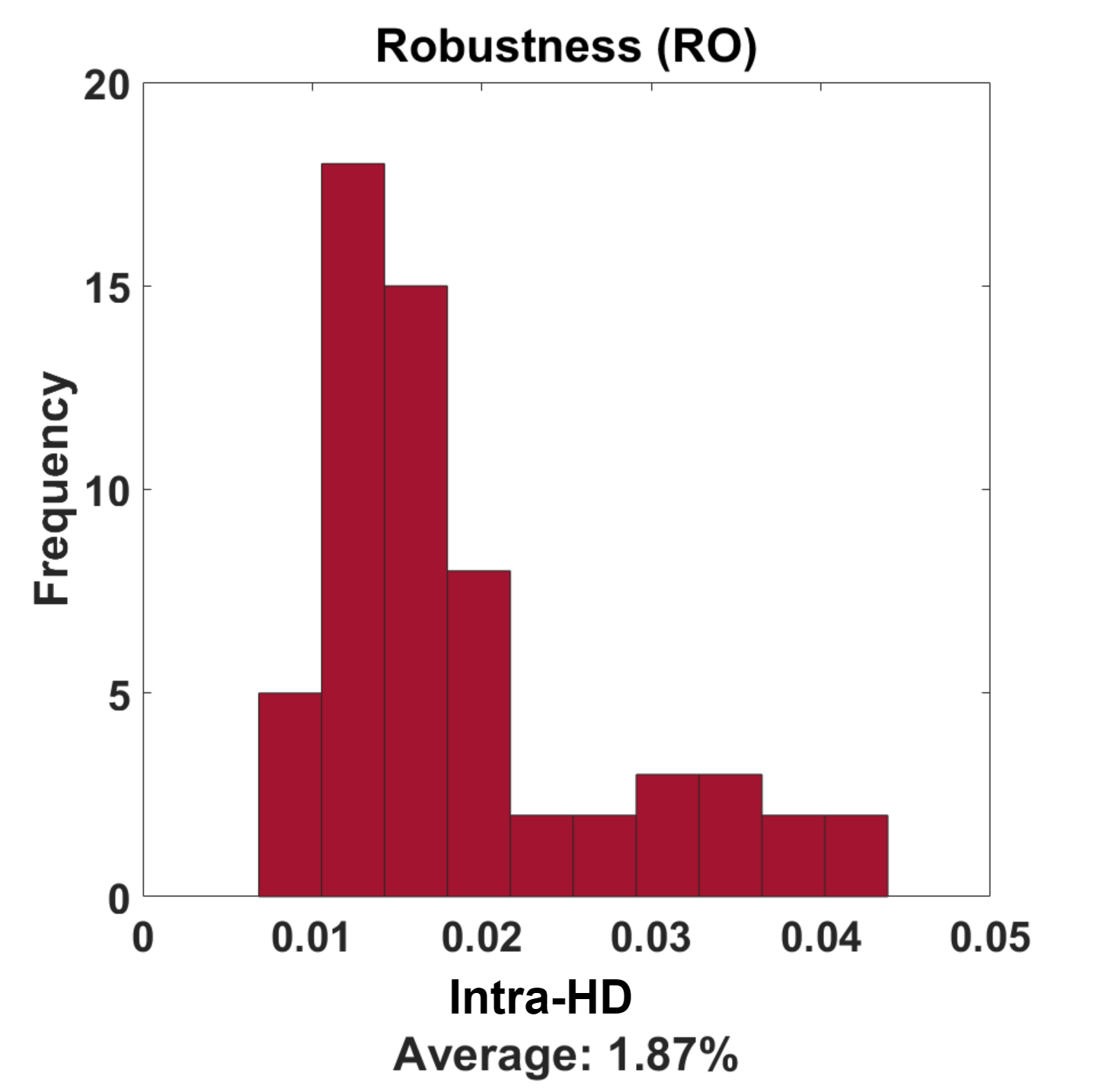}
                \label{fig:rointra}
        }
        
        \centering
       \caption{Comparison of (a) uniqueness; and (b) robustness between \melpuf, SRAM PUF and RO PUF (at $T_{NOM} = 25$  $^\circ$C and $V_{supply,NOM} = 3.3$  V). We generate the results for \melpuf~~structure using the D4 distribution as shown in Fig.~\ref{fig:h9}.}
       \label{fig:resutsCompPUF}
\end{figure*}

 %%%%%%%%%%%%%%%%%
\begin{table*}
  \centering
  \scriptsize
  \caption{Comparison of NIST Test Suite randomness results  between SRAM PUF, RO PUF, and MeLPUF for 1 million response bits.}
\label{tab:NISTComp}
    \begin{tabular}{|c|c|c|c|c|c|c|}
    \hline
    \multirow{2}[4]{*}{\textbf{Test Name}} & \multicolumn{2}{c|}{\textbf{SRAM PUF}} & \multicolumn{2}{c|}{\textbf{RO PUF}} & \multicolumn{2}{c|}{\textbf{MeLPUF}} \\
\cline{2-7}          & \multicolumn{1}{c|}{\begin{tabular}[c]{@{}c@{}}No. of\\ Passing\\ Inputs\end{tabular}} & \multicolumn{1}{c|}{\textit{p}-value} & \multicolumn{1}{c|}{\begin{tabular}[c]{@{}c@{}}No. of\\ Passing\\ Inputs\end{tabular}} & \multicolumn{1}{c|}{\textit{p}-value} & \multicolumn{1}{c|}{\begin{tabular}[c]{@{}c@{}}No. of\\ Passing\\ Inputs\end{tabular}} & \multicolumn{1}{c|}{\textit{p}-value} \\
   [0.5ex] \hline
    \textbf{Frequency} & 10/10 & 0.21331 & 2/10  & 0.00000 & 10/10 & 0.73992 \\
    [0.5ex]\hline
    \textbf{Block Frequency} & 10/10 & 0.06688 & 0/10  & 0.00000 & 10/10 & 0.35049 \\
    [0.5ex]\hline
    \textbf{Cummulative Sums, $S_1$} & 10/10 & 0.73992 & 0/10  & 0.00000 & 10/10 & 0.91141 \\
    [0.5ex]\hline
    \textbf{Cummulative Sums, $S_2$} & 10/10 & 0.73992 & 1/10  & 0.00000 & 10/10 & 0.73992 \\
    [0.5ex]\hline
    \textbf{Runs} & 10/10 & 0.12233 & 1/10  & 0.00000 & 10/10 & 0.03517 \\
    [0.5ex]\hline
    \textbf{Longest Runs} & 10/10 & 0.12233 & 2/10  & 0.00000 & 10/10 & 0.73992 \\
    [0.5ex]\hline
    \textbf{Rank} & 10/10 & 0.00095 & 10/10  & 0.00000 & 10/10 & 0.00020 \\
    [0.5ex]\hline
    \textbf{FTT} & 10/10 & 0.35049 & 9/10  & 0.00020 & 10/10  & 0.21331 \\
    [0.5ex]\hline
    \end{tabular}%
%   \label{tab:addlabel}%

\vspace{-10pt}
\end{table*}%

%%%%%%%%%%%%%%%%%

%%%%%%%%%
\begin{table}
\scriptsize
\caption{Comparison of the uniqueness and robustness between \melpuf~~ and the related works introduced in Section~\ref{sec:intro}.}
%\textcolor{black}{chris please check the caption once}}
\label{tab:compprev}
\centering
\begin{tabular}{|c|c|c|c|c|}
\hline
\begin{tabular}[c]{@{}c@{}}\textbf{{PUF}}\\\textbf{configuration} \end{tabular}  & \begin{tabular}[c]{@{}c@{}}\textbf{Inter-HD} \end{tabular} & \begin{tabular}[c]{@{}c@{}}\textbf{Intra-HD}\end{tabular}& 
\begin{tabular}[c]{@{}c@{}}\textbf{ML Attack}\\\textbf{(Resistance)} \end{tabular} &
\begin{tabular}[c]{@{}c@{}}\textbf{Synthesizability} \end{tabular} \\ \hline

\textbf{LUT} \cite{bib:gan} &49.07\% & 1.20\% &\xmark&\xmark\\ \hline
\textbf{LUTSR} \cite{bib:wang} & 47.10\% & 1.00\% &\xmark&\xmark\\ \hline
\textbf{LFSR} \cite{bib:ams} & 47.40\% & 3.30\% &\xmark&\xmark\\ \hline
\textbf{XRBR} \cite{bib:cui} (a) & 40.67\% & 1.78\% &\cmark&\xmark\\ \hline
\textbf{XRRO} \cite{bib:cui} (b) & 48.76\% & 2.28\% &\cmark&\xmark\\ \hline
\textbf{SCAN} \cite{bib:scan} & 47.14\% & 3.17\% &\xmark&\xmark\\ \hline
\textbf{MECCA} \cite{bib:meca} & 49.90\% & 0.85\% &\xmark&\xmark\\ \hline
\textbf{MeLPUF} & 50.05\% & 2.57\% &\cmark&\cmark\\ \hline
\end{tabular}
% \vspace{-10pt}
\end{table}
%%%%%%%%%

%%%%%%%%%
% \begin{comment}

\begin{table}
\scriptsize
\caption{Overhead comparison between an RO, SRAM, and \melpuf~for a 1024-bit response.}
\label{tab:comp}
\centering
\begin{tabular}{|c|c|c|c|c|c|}
\hline

\begin{tabular}[c]{@{}c@{}}\textbf{{PUF}}\\\textbf{configuration} \end{tabular}  & \begin{tabular}[c]{@{}c@{}}\textbf{Area}\\\textbf{(LE)} \end{tabular} & \begin{tabular}[c]{@{}c@{}}\textbf{Power}\\\textbf{(mW)} \end{tabular}& \begin{tabular}[c]{@{}c@{}}\textbf{Delay}\\\textbf{(ns)} \end{tabular} & \textbf{Inter-HD} & \textbf{Intra-HD} \\ \hline
\textbf{Baseline} & 1,650 & 255.94 & 11.40 & N/A & N/A\\ \hline
\textbf{RO PUF} & 16,983 & 295.70 & 14.42 & 42.53\% & 1.87\%\\ \hline
\textbf{SRAM PUF} & 5,977 & 294.44 & 12.25 & 50.50\% & 2.13\%\\ \hline
\textbf{\melpuf~} & 4,743 & 275.81 & 13.39 & 50.05\% & 2.57\% \\ \hline
\end{tabular}
\vspace{-10pt}
\end{table}
% \end{comment}
%%%%%%%%%
\subsection{Comparison with Existing PUFs} 
\label{sec:comp}

To ensure an unbiased comparison with \melpuf, we implement an RO PUF and SRAM PUF on the same platform as \melpuf. \textcolor{black}{We also use the same experimental set up as discussed in Section~\ref{sec:results}.} We select a signature length of 1024-bits to compare the quality metrics: uniqueness, robustness, and randomness. 

We use the HaHa board, which contains an SRAM module for implementing the SRAM PUF. 
%The module is controlled by an ATmega microcontroller that transmits data through interconnects to the FPGA. 
During power-up, the SRAM values are transmitted 1 Byte at a time and stored in the FPGA. It requires supplemental memory and logic elements to read out the signatures. It results in a marginally higher area overhead in terms of logic elements and hardware overhead. Fig.~\ref{fig:sraminter} and~\ref{fig:sramintra} elucidate the inter and intra-Hamming distances respectively. Table \ref{tab:NISTComp} demonstrates the randomness results for the NIST test suite. The results collected from this implementation show that \melpuf~~ is comparable to an SRAM PUF with a significantly reduced overhead. However, \melpuf~~shows a significantly higher level of reliability as shown in Fig. \ref{tab:volt}. This shows that that \melpuf~ is significantly more resilient to voltage variations than SRAM PUFs.

We also implement an RO PUF on the HaHa board. We ensure that the symmetry between the ROs is maintained by confining the RO PUF to specific regions of the FPGA ~\cite{bib:LABsym}. This restriction limits the reconfigurability of the RO PUF, unlike \melpuf, which is easily reconfigurable. The structure of the RO PUF also increases the used area significantly, occupying three times the area compared to \melpuf~~and the SRAM PUF. 

Fig. \ref{fig:sraminter} and Fig. \ref{fig:sramintra} show the uniqueness and robustness results for the three PUF implementations. We examine that \melpuf~~performs similarly or better in both uniqueness and randomness, with comparable results in terms of robustness. Table \ref{tab:NISTComp} shows the randomness results for the NIST test suite. The randomness of the RO PUF implementation is very poor compared to both \melpuf~~and the SRAM PUF but has very high reliability. This high reliability can be explained by the RO PUF not requiring a mirrored symmetrical connection for its PUF values ~\cite{bib:morrorRO}. The lack of this requirement allows it to mitigate the negative effects in terms of robustness caused by the FPGA architecture. 

%%%%%%%%%
\begin{comment}

\begin{table}[]
\scriptsize
\caption{Comparison of overhead results between the implemented RO, SRAM, and \melpuf.\textcolor{black}{}}
\label{tab:comp}
\centering
\begin{tabular}{|c|c|c|c|c|c|}
\hline

\begin{tabular}[c]{@{}c@{}}\textbf{{PUF}}\\\textbf{configuration} \end{tabular}  & \begin{tabular}[c]{@{}c@{}}\textbf{Area}\\\textbf{(LE)} \end{tabular} & \begin{tabular}[c]{@{}c@{}}\textbf{Power}\\\textbf{(mW)} \end{tabular}& \begin{tabular}[c]{@{}c@{}}\textbf{Delay}\\\textbf{(ns)} \end{tabular} & \textbf{Inter-HD} & \textbf{Intra-HD} \\ \hline
\textbf{No PUF} & 1,650 & 255.94 & 11.40 & N/A & N/A\\ \hline
\textbf{RO PUF} & 16,983 & 295.70 & 14.42 & 42.53\% & 1.87\%\\ \hline
\textbf{SRAM PUF} & 5,977 & 294.44 & 12.25 & 50.50\% & 2.13\%\\ \hline
\textbf{\melpuf~} & 4,743 & 275.81 & 10.29 & 47.30\% & 3.10\% \\ \hline
\end{tabular}
\vspace{-10pt}
\end{table}
\end{comment}
%%%%%%%%%

In order to achieve a stable PUF response the PUF elements mush be distributed throughout a circuit. We inserted \melpuf~elements into the ISCAS circuit and compared the incurred overhead to the same circuit utilizing with the SRAM and RO PUFs. Table \ref{tab:comp} compares the area, power, and delay overheads, along with the inter-HD and intra-HD, between the test circuit without a PUF, an SRAM PUF, an RO PUF, and \melpuf. Each PUF was configured to add 1024 PUF elements which are used to collect a 1024-bit response. For \melpuf, this involves modifying 1024 gates present in the circuit. While for the SRAM and RO PUF, 1024 PUF elements were added along side the circuit. We observe that our proposed structure incurs less overhead for all three parameters than an RO PUF or an SRAM PUF for the same signature length. Additionally, Table \ref{tab:compprev} shows a comparison of the uniqueness and robustness for previous works along with \melpuf. We inspect that MeLPUF surpasses the performance of others when compared to the existing works.

	%%%%%%%%%% Conclusion  %%%%%%%%%%
	\section{Conclusion}
\label{sec:conclusion}

We have presented \melpuf, a synthesizable and distributed memory-based PUF structure, which can be inserted into an existing logic circuit for authentication or cryptographic key generation. We demonstrate that \melpuf~enables the designer to seamlessly integrate the PUF structures at higher design abstractions leading to lower overheads. We evaluate the \melpuf~designs using simulation and hardware evaluations on Intel MAX 10 FPGA devices and show that \melpuf~exhibits the uniqueness, robustness and reliability properties of an ideal PUF. We also demonstrate that \melpuf~incurs lower overheads, and has better attack resistance when compared other state-of-the-art structures. In future, we plan on investigating the feasibility of other PUF structures and the scalalability of \melpuf~to complex designs such as System-on-Chip architectures.

%It enables a designer to transform select logic gates into a bi-stable memory element, whose power-up states act as the entropy source. It also enables us to create a decentralized structure that can be integrated throughout any complex design. We have presented a methodology on utilizing this feature by synthesizing \melpuf~~ into existing circuit designs. We have extensively evaluated the PUF through circuit-level simulations using HSPICE and experimental measurements on Intel MAX 10 FPGA devices. The experimental results have demonstrated near-ideal uniqueness and robustness of the generated signatures. A judicious selection of LUTs has shown an average uniqueness of 50.05\%and robustness of 2.57\%. Further verification regarding supply voltage variations also exhibited stability of the generated signatures at different voltage levels. \melpuf~~ has also demonstrated superior performance in terms of uniqueness, robustness, and area overhead compared to alternative PUF techniques. We have also investigated different attack scenarios and demonstrated that \melpuf~~can be configured to resist various potential attacks. Model building attacks are inapplicable, while side-channel attacks can be mitigated through careful integration of the PUF into logic circuits. 

\bibliographystyle{IEEEtran}
\bibliography{samples/sample-base.bib}

\end{document}